\documentclass[12pt]{article}
\usepackage{graphicx}

\font\grande=cmr9.5 scaled \magstep4
\font\medio=cmr9.5 scaled \magstep2
\outer\def\beginsection#1\par{\medbreak\bigskip
      \message{#1}\leftline{\bf#1}\nobreak\medskip
\vskip-\parskip
      \noindent}

\begin{document}
\bibliographystyle {unsrt}

\titlepage

\begin{flushright}
CERN-PH-TH/2009-174
\end{flushright}

\vspace{15mm}
\begin{center}
{\grande Circular dichroism, magnetic knots}\\ 
\vskip 6mm
{\grande and the spectropolarimetry}\\
\vskip 6mm
{\grande of the Cosmic Microwave Background}
\vskip1.cm
 
Massimo Giovannini\footnote{Electronic address: massimo.giovannini@cern.ch}

\vskip1cm
{\sl Department of Physics, Theory Division, CERN, 1211 Geneva 23, Switzerland}
\vskip 0.2cm
{\sl INFN, Section of Milan-Bicocca, 20126 Milan, Italy}
\end{center}
\vskip 2cm
\centerline{\medio  Abstract}
When the last electron-photon scattering takes place in a magnetized environment, the degree of circular polarization 
of the outgoing radiation depends upon the magnetic field strength. 
After deriving the scattering matrix of the process, the generalized radiative transfer equations 
are deduced in the presence of the relativistic fluctuations of the geometry and for all the four brightness perturbations.
The new  system of equations is solved under the assumption that the incident radiation is not polarized. 
The induced V-mode polarization is analyzed both analytically and numerically. 
The corresponding angular power spectra are calculated 
and compared with the measured (or purported) values of the linear polarizations (i.e. E-mode and B-mode) as they arise in the concordance model and in its neighboring extensions. Possible 
connections between the V-mode polarization of the Cosmic Microwave background 
and the topological properties of the magnetic flux 
lines prior to equality are outlined and briefly explored in analogy with the physics of magnetized sun spots.  
\noindent

\vspace{5mm}

\vfill
\newpage
\renewcommand{\theequation}{1.\arabic{equation}}
\setcounter{equation}{0}
\section{Motivations and goals}
\label{sec1}
The circular polarization of the Cosmic Microwave Background (CMB in what follows) is not the direct target 
of forthcoming experimental searches. It will be argued hereunder that more accurate spectropolarimetric 
measurements of the V-mode polarization can be enlightening especially as a diagnostic of the 
magnetization of the pre-decoupling plasma. 
The primary goal of experimental endeavors in the near future is related, in one way or in the 
other, to the determination of the angular power spectra of the intensity and of the linear polarization of the CMB
radiation field. Even the B-mode polarization, one of the primary objectives of diverse experimental programs, 
will be unable to shed light on the circular polarizations of the CMB. To avoid possible misunderstandings on this 
point it is desirable to introduce the relevant conventions on the Stokes parameters of the radiation field
\footnote{From the definitions of the Stokes parameters  it follows  
(see, e. g. \cite{jack,chandra,pera}) that $I^2 \geq Q^2 + U^2 + V^2$, where the equality sign 
arises in the case of the field of a plane wave.}
\begin{eqnarray}
&& I = |\vec{E}\cdot\hat{e}_{1}|^2 +  |\vec{E}\cdot\hat{e}_{2}|^2,\qquad V = 2 \,\mathrm{Im}[ (\vec{E}\cdot\hat{e}_{1})^{*} (\vec{E}\cdot\hat{e}_{2})],
\label{IV}\\
&& Q = |\vec{E}\cdot\hat{e}_{1}|^2 -  |\vec{E}\cdot\hat{e}_{2}|^2, \qquad U = 2 \,\mathrm{Re}[ (\vec{E}\cdot\hat{e}_{1})^{*} (\vec{E}\cdot\hat{e}_{2})],
\label{QU}
\end{eqnarray}
where $\hat{e}_{1}$ and $\hat{e}_{2}$ are two mutually orthogonal directions both perpendicular  
to the direction of propagation $\hat{n}$ which is chosen to lie along $\hat{e}_{3}$.  
The temperature autocorrelation 
(i.e. the TT angular power spectrum\footnote{The specific definitions of the relevant angular power spectra are provided hereunder, see, e.g. Eqs. (\ref{APS5}) and (\ref{APS6})}) stems directly from the 
brightness perturbation of the intensity of the radiation field: the fluctuations 
of the space-time curvature act as sources of inhomogeneity of the intensity. The 
treatment of the polarization is slightly more cumbersome and it has to do with the transformation 
properties of the Stokes parameters. Consider a rotation of the unit vectors $\hat{e}_{1}$ and $\hat{e}_{2}$ 
in the plane orthogonal to $\hat{n}$ and suppose that the rotation angle is $\varphi$. By rotating the axes in the 
right-handed sense, i.e. as 
\begin{equation}
\hat{e}_{1}' = \cos{\varphi}\,\hat{e}_{1} + \sin{\varphi}\,\hat{e}_{2},\qquad 
\hat{e}_{1}' = -\sin{\varphi}\,\hat{e}_{1} + \cos{\varphi}\,\hat{e}_{2},
\label{rotation1}
\end{equation}
the Stokes parameters of Eq. (\ref{IV}) are left invariant (i.e. $I' = I$ and $V'=V$) 
while $Q$ and $U$ (introduced in Eq. (\ref{QU})) transform as
\begin{equation}
 Q' = \cos{2\varphi}\,Q + \sin{2\varphi}\,U,\qquad  U' = -\sin{2 \varphi}\,Q + \cos{2\varphi}\,U.
\label{rotation2}
\end{equation}
In different words, while $I$ and $V$ are invariant under a two-dimensional rotation, $Q$ and $U$ 
do transform and do mix under the same rotation. It is worth stressing that he transformation parametrized by Eq. (\ref{rotation1}) 
is not a global rotation of the coordinate system but it is rather a rotation on the tangent plane to the two-sphere 
at a given point.  
The transformation properties of the Stokes parameters under Eq. (\ref{rotation1}) allow determine
their associated spin weight \cite{sh0,sh1,sh2,sh3}. The brightness perturbations associated with $I$ and $V$ 
(i.e. $\Delta_{\mathrm{I}}$ and $\Delta_{\mathrm{V}}$) have both spin-weight $0$. The brightness perturbations 
of $Q$ and $U$ (i.e. $\Delta_{Q}$ and $\Delta_{U}$) transform as a function of spin-weight $\pm2$, since, 
from Eq. (\ref{rotation2}), 
$\Delta_{\pm}(\hat{n},\tau) = \Delta_{\mathrm{Q}}(\hat{n},\tau) \pm i \Delta_{\mathrm{U}}(\hat{n},\tau)$ 
transform as, $\Delta^{\prime}_{\pm}(\hat{n},\tau) = e^{\mp 2 i \varphi} \Delta_{\pm}(\hat{n},\tau)$.
Consequently, while $\Delta_{\pm}(\hat{n},\tau)$ have to be expanded, on the two-sphere,  in terms of spin $\pm 2$ spherical harmonics $_{\pm 2}Y_{\ell\, m}(\hat{n})$
\begin{equation}
\Delta_{\pm}(\hat{n},\tau) = \sum_{\ell \, m} a_{\pm\,2,\,\,\ell\, m} \, _{\pm 2}Y_{\ell\, m}(\hat{n}),
\label{sp2}
\end{equation}
$\Delta_{\mathrm{I}}(\hat{n},\tau)$ and $\Delta_{\mathrm{V}}(\hat{n},\tau)$ 
have to be  expanded on the basis of spherical harmonics $Y_{\ell\,m}(\hat{n})$ as:
\begin{equation}
\Delta_{\mathrm{I}}(\hat{n}, \tau) = \sum_{\ell m} a_{\ell\, m}^{(\mathrm{I})} \, Y_{\ell\, m}(\hat{n}), \qquad 
\Delta_{\mathrm{V}}(\hat{n}, \tau) = \sum_{\ell m} a_{\ell\, m}^{(\mathrm{V})} \, Y_{\ell\, m}(\hat{n}).
\label{sp0}
\end{equation}
Both spin-0 and spin-$\pm 2$ spherical harmonics arise naturally as Wigner matrix elements \cite{sakurai} depending 
in general upon three different eigenvalues: while the third eigenvalue is $0$ for scalar harmonics, it is $\mp s$ 
for spin-$s$ weighted harmonics.
The complete information on the radiation field of the CMB should therefore stem from 
the analysis of the TT and VV angular power spectra
\begin{equation}
C_{\ell}^{(\mathrm{TT})} = \frac{1}{2\ell +1} \sum_{m = - \ell}^{\ell} \langle a_{\ell\, m}^{(\mathrm{I})*} 
a_{\ell\, m}^{(\mathrm{I})}\rangle, \qquad 
C_{\ell}^{(\mathrm{VV})} = \frac{1}{2\ell +1} \sum_{m = - \ell}^{\ell} \langle a_{\ell\, m}^{(\mathrm{V})*} 
a_{\ell\, m}^{(\mathrm{V})}\rangle,
\label{APS1}
\end{equation}
as well as from the angular power spectra of the E-mode and B-mode autocorrelations \footnote{This statement on the autocorrelations 
of temperature and polarization does not exclude the possibility of discussing and measuring 
the various cross-correlations between temperature and polarization.}. The 
E-mode and B-mode autocorrelations are defined as
\begin{equation}
C_{\ell}^{(\mathrm{EE})} = \frac{1}{2\ell + 1} \sum_{m = -\ell}^{\ell} 
\langle a^{(\mathrm{E})*}_{\ell m}\,a^{(\mathrm{E})}_{\ell m}\rangle,\qquad 
C_{\ell}^{(\mathrm{BB})} = \frac{1}{2\ell + 1} \sum_{m=-\ell}^{\ell} 
\langle a^{(\mathrm{B})*}_{\ell m}\,a^{(\mathrm{B})}_{\ell m}\rangle,
\label{APS2}
\end{equation}
where $a^{(\mathrm{E})}_{\ell\, m}$ and $a^{(\mathrm{B})}_{\ell\, m}$ are 
a linear combination of the coefficients $a_{\pm\,2,\,\,\ell\, m}$  already introduced in Eq. (\ref{sp2}):
\begin{equation}
a^{(\mathrm{E})}_{\ell\, m} = - \frac{1}{2}(a_{2,\,\ell m} + a_{-2,\,\ell m}), \qquad  
a^{(\mathrm{B})}_{\ell\, m} =  \frac{i}{2} (a_{2,\,\ell m} - a_{-2,\,\ell m}).
\label{APS3}
\end{equation}
In real space the fluctuations constructed from 
$a^{(\mathrm{E})}_{\ell\,m}$ and $a^{(\mathrm{B})}_{\ell\,m}$ have the 
property of being invariant under rotations on a plane orthogonal 
to $\hat{n}$ and, up to an $\ell$-dependent prefactor, they can 
be expanded in terms of (ordinary) spherical harmonics:
\begin{equation}
\Delta_{\mathrm{E}}(\hat{n},\tau) = \sum_{\ell\, m} N_{\ell}^{-1} \,  a^{(\mathrm{E})}_{\ell\, m}  \, Y_{\ell\, m}(\hat{n}),\qquad 
\Delta_{\mathrm{B}}(\hat{n},\tau) = \sum_{\ell\, m} N_{\ell}^{-1} \,  a^{(\mathrm{B})}_{\ell\, m}  \, Y_{\ell\, m}(\hat{n}),
\label{APS4}
\end{equation}
where $N_{\ell} = \sqrt{(\ell - 2)!/(\ell +2)!}$.  Before matter radiation equality the radiation field is customarily 
assumed to be unpolarized. The properties 
of electron-photon (and ion-photon) scattering imply that the radiation can become linearly polarized 
provided the incident brightness perturbations have a non-vanishing quadrupole moment\footnote{As usual 
$\Delta_{\mathrm{I}\ell}$ denotes the $\ell$-th multipole of the intensity of the radiation field. With 
the same notation we will refer, when needed, to the multipoles of other observables.}, i.e.
$\Delta_{\mathrm{I}2} \neq 0$. 

Circular dichroism as well as linear polarization
of the CMB becomes theoretically plausible in the presence of pre-equality magnetic fields \cite{rev1,rev2} but, 
so far, there has not been any specific calculation of the circular polarization induced by a magnetized plasma 
prior to recombination and in the framework of the concordance model. One of the purposes of the present 
article is to fill such a gap. While the V-mode polarization is suppressed in comparison 
with the E-mode polarization, the question is to determine 
quantitatively the nature of the suppression and its typical range in multipole 
space.  The answer to the latter question can only be dynamical: it will be interesting, for the 
present purposes, to understand how, when and to what extent a radiation field 
which is originally unpolarized prior to equality will become circularly polarized after 
photon decoupling. 

Some of the phenomenological aspects of the present considerations can be understood 
in analogy with solar spectropolarimetry. The knowledge of large-scale magnetism is often inspired (and partially 
modeled) by our improved understanding of solar magnetism (see, for instance, \cite{suns1} for a dedicated review on the subject). Circular polarization is known to occur in sunspots \cite{suns2,suns3,suns4}  and it is sometimes argued that the analysis of polarization might improve the understanding of the local topology of magnetic flux lines 
in the vicinity of sunspots \cite{suns1}.  The amount of circular polarization depends, in the case of the 
sun, both on the gradients of the velocity field and as well as upon the magnetic field topology and intensity. 
A naive approach to the problem would suggest that, absent any velocity gradient,
$V \simeq 9\times 10^{-12} B_{\mathrm{spot}} \,\, \lambda\,\, I$ where $\lambda$ is the wavelength in units of 
nm ($1 \mathrm{nm} = 10^{-9} \, \mathrm{m}$) and $B_{\mathrm{spot}}$ is the magnetic field in units of gauss. As noted 
long ago (see e. g. \cite{suns2,suns3})
for typical values of the magnetic field (i.e. $B_{\mathrm{spot}} \sim \mathrm{kG}$) and for optical wavelengths (i.e. 
$380\, \mathrm{nm} < \lambda < 750 \,\mathrm{nm}$) we would get  $V/I \sim 10^{-5}$  which is one or two orders of magnitude smaller 
than what approximately observed. The latter discrepancy suggested (see e.g. \cite{suns5})  the important role played by velocity gradients 
whose contribution should be included for a consistent interpretation of the observational data. 

Circular polarization arises naturally also in synchrotron emission \cite{sync1,sync2,sync2a} (see 
also \cite{sync3,sync4}). In this second context the amount  of circular polarization is generated from the 
linear polarization by the so called Faraday conversion effect.  In the case of the CMB the relativistic 
effect associated with the synchrotron emission is difficult to realize. On the contrary
the pre-decoupling plasma is rather cold and the electrons are non relativistic. Furthermore, 
it is difficult to justify that the initial radiation field should be linearly polarized already well before equality.
The only way Faraday conversion could lead to circular polarization is as a secondary effect 
when the  linearly polarized  CMB photons impinge on the relativistic electrons in the 
cluster magnetic field \cite{sync5}. In this paper, on the contrary, the target is to compute
the amount of circular polarization induced at early times by a pre-decoupling magnetic field and in terms 
of unpolarized initial conditions of the Einstein-Boltzmann hierarchy. 
To introduce some quantitative (but still general) considerations it is practical to define, for immediate convenience, 
\begin{eqnarray}
G_{\ell}^{(\mathrm{TT})} &=&
 \frac{\ell (\ell +1)}{2 \pi} C_{\ell}^{(\mathrm{TT})}, \qquad 
 G_{\ell}^{(\mathrm{VV})} = \frac{\ell (\ell +1)}{2 \pi} C_{\ell}^{(\mathrm{VV})},
\label{APS5}\\
G_{\ell}^{(\mathrm{EE})} &=&
 \frac{\ell (\ell +1)}{2 \pi} C_{\ell}^{(\mathrm{EE})}, \qquad 
 G_{\ell}^{(\mathrm{BB})} = \frac{\ell (\ell +1)}{2 \pi} C_{\ell}^{(\mathrm{BB})},
\label{APS6}
\end{eqnarray}
measuring, for a given observable, the angular power per logarithmic interval of $\ell$.
With the same notation the cross-correlations between different observables can be 
defined, for instance, as $G_{\ell}^{(\mathrm{TE})}$, $G_{\ell}^{(\mathrm{VT})}$ and so on and so forth.
In terms of Eqs. (\ref{APS5}) and (\ref{APS6}) the typical orders of magnitude 
angular power spectra can then be be summarized as \footnote{For sake of simplicity the numerical values 
quoted here refer to the maximum of each power spectrum}
\begin{equation}
G_{\ell}^{(\mathrm{TT})} \simeq {\mathcal O}(5\times 10^{3}) (\mu \, \mathrm{K})^2, \qquad 
G_{\ell}^{(\mathrm{TE})} \simeq {\mathcal O}(150) (\mu \, \mathrm{K})^2,\qquad 
G_{\ell}^{(\mathrm{EE})} \simeq {\mathcal O}(50) (\mu \, \mathrm{K})^2.
\label{est1}
\end{equation}
The TT and TE correlations have been accurately assessed by the WMAP 
collaboration \cite{WMAP5a, WMAP5b,WMAP5c} (see also \cite{WMAP3a,WMAP3b}). Interesting measurements on the EE 
correlations have been reported, for instance,  by the QUAD experiment \cite{QUAD1,QUAD2,QUAD3,QUAD4}.  
Other measurements on the TT correlation for large multipoles (i.e. $\ell > 1000$)  have been 
reported by the ACBAR experiment \cite{ACBAR1,ACBAR2}. While all the current
experimental data are consistent with the standard $\Lambda$CDM paradigm they can also be used 
to estimate the parameters of a putative magnetized background. In this respect, the result is 
that large-scale magnetic fields of nG strength and slightly blue spectral indices 
are allowed by current CMB data \cite{max1,max2}.  The estimation of the parameters 
of a magnetized background led to the first estimate of the likelihood contours in the 
two-dimensional plane characterized by the magnetic spectral index and by the 
magnetic field intensity. In a frequentist perspective the results of \cite{max1,max2} exclude, to 
95\% confidence level, a sizable portion of the parameter space of magnetized models
centered around comoving field intensities of $3.2$ nG and magnetic spectral indices\footnote{The conventions 
on the magnetic spectral indices $n_{\mathrm{B}}$ are exactly the same as the ones 
employed for the spectral index of curvature perturbations (denoted as $n_{\mathrm{s}}$). 
The scale-invariant value corresponds, in both cases, to $1$.} $n_{\mathrm{B}} = 1.6$.
The latter results hold when the underlying model is just the $\Lambda$CDM (concordance) paradigm\footnote{In the acronym 
$\Lambda$ stands for the (non-fluctuating) dark energy component while CDM stands for cold dark matter.}
but the addition of, for instance, dark energy fluctuations does not change quantitatively the 
exclusion plots \cite{max2a}. 

The B-mode polarization has not been measured yet and upper limits exist by various experiments. 
In the standard $\Lambda$CDM paradigm with no tensors the BB power spectrum vanishes.
For instance magnetic fields of ${\mathcal O}(5\, \mathrm{nG })$ 
and blue spectral index lead, at intermediate scales, to an angular power 
spectrum which can be as large as $10^{-3}\,\, (\mu \mathrm{K})^2$ \cite{max3}. The latter estimate
can be compared, for instance, with the BB angular power spectrum expected 
from the tensor modes of the geometry and from the gravitational lensing of the CMB anisotropies. 
The tensor modes are the conventional (potential) source of  B-mode polarization 
in the simplest extension of the $\Lambda$CDM paradigm. Defining $r_{\mathrm{T}}$ as the tensor to scalar 
ratio, a typical value $r_{\mathrm{T}} \sim 0.3$ would imply, at intermediate 
multipoles (i.e. $\ell < 100$) a BB angular power spectrum of the order of $10^{-2} \,\, (\mu \mathrm{K})^2$. 

The degree of circular polarization computed here 
depends, in the simplest case, upon the amount of curvature perturbations, upon the magnetic field parameters and upon the 
typical frequency of the experiment. For nG magnetic fields and for a reference frequency
${\mathcal O}(10)\, \mathrm{GHz}$ the VV angular power spectrum is $G_{\ell}^{(\mathrm{VV})} \simeq 10^{-15} (\mu \mathrm{K})^2$.
For the same range of parameters the cross-correlation 
 $G_{\ell}^{(\mathrm{VT})} \simeq 10^{-6.5} (\mu \mathrm{K})^2$. If the observational frequency decreases the 
 signals can be larger.  While this comparison will be more 
 carefully performed in section 4 it is important to appreciate that, in some sense, the absolute magnitude 
 of the different correlation functions represents just a necessary but insufficient guide for 
 the observer since the systematics associated with the circular polarization 
 are different from the ones arising in the case of  a linearly polarized signal \cite{gs0}.
The  second point we wish to stress is that, as it will be apparent from the subsequent analysis, 
 low frequency instruments seem to be experimentally preferable \cite{LF1,LF2} and, in this respect, 
 it is tempting to speculate that very low frequency (radio) techniques could be 
 appropriately adapted \cite{gs1,gs2,gs3}. The third point related to the potential observations 
 of the effects is that spectropolarimetric techniques should be probably employed given 
 the necessity of a simultaneous determination of the brightness perturbation of the intensity and of the 
 V-mode polarization (see, in this respect, section \ref{sec4}).

The analogy of the present problem with the physics of the sunspots suggests the possibility of connecting the 
amount of (measured) circular polarization with the topology of the magnetic flux lines. This 
topic is, in principle rather rich and, in this paper, we will merely scratch the surface by presenting some particular examples which are only semi-realistic and which are borrowed from known examples 
in plasma physics. Indeed, the study of the topology of magnetic flux lines has a long history going back to the 
pioneering work of Fermi and Chandrasekhar on the gravitational stability of the galactic arm \cite{fermi1,fermi2}.
It would be interesting, in perspective, to connect the possible occurrence (or absence) of the circular polarization 
with magnetized plasmas which  minimize the magnetic energy while the helicity 
is conserved (as it should) at high conductivity (see, along this line, the seminal papers of Chandrasekhar, Kendall 
and Woltjer \cite{ch1,ch2,wolt}).

The layout of the paper is the following. In section \ref{sec2} the photon electron and photon ion 
scattering will be discussed in the presence of a magnetic field in the guiding centre approximation. Details 
are also reported in appendix \ref{APPA}. In section \ref{sec3} the same problem will be addressed in the case 
of a magnetic knot, i.e. a simple example of static configuration minimizing 
the energy at a fixed value of the magnetic helicity. It will be speculated that the degree 
of circular polarization could be eventually connected with the topological properties of the magnetic flux lines in the plasma.
In section \ref{sec4} the evolution equations of the brightness 
perturbations will be deduced and solved both analytically and numerically. Relevant details on this topic are 
given in appendix \ref{APPB}. The tight-coupling approximation 
will be applied to the new framework and analytical results for the V-mode power spectra for large angular scales will be derived. 
For smaller angular scales numerical results will also be presented and compared with the temperature and 
(linear) polarization anisotropies. Section \ref{sec5} contains the concluding considerations. 

\renewcommand{\theequation}{2.\arabic{equation}}
\setcounter{equation}{0}
\section{Magnetized electron-photon scattering}
\label{sec2}
The electron-photon scattering is customarily computed without taking into account 
the contribution of the magnetic field itself to the scattering matrix. This happens not only 
in the case when magnetic fields are assumed to be absent but also in the presence 
of large-scale magnetic fields (see, for instance\footnote{It is appropriate to stress that it 
would be rather pretentious to give complete and thorough list of references in connection with primordial magnetism. 
The easiest solution is to refer the interested reader to the dedicated review articles of \cite{rev1,rev2} 
where a more complete bibliography can be found.}, \cite{add1,add2,add3}). The purpose of this 
section is to drop such an assumption and to derive the appropriate scattering matrix for electron-photon 
scattering in a weakly magnetized plasma. 
It is practical to define, for the present purposes, the outgoing and ingoing Stokes vectors  whose 
components are the Stokes parameters, i.e. 
\begin{eqnarray}
&& {\mathcal I}^{\mathrm{out}}(\omega,\,\mu,\,\mu',\,\varphi,\, \varphi') = (I_{1},\, I_{2},\, U,\, V),
\label{OUT}\\
&&  {\mathcal I}^{\mathrm{in}}(\omega,\,\mu',\,\varphi') = (I_{1}',\, I_{2}',\, U',\, V'),
\label{IN}
\end{eqnarray}
 $\mu = \cos{\vartheta}$ and $\mu' = \cos{\vartheta'}$. 
The intensity $I$ and one of the components of the linear polarization (i.e. $Q$) have been 
replaced, as usual,  by $I_{1} = (I+ Q)/2$ and $I_{2}=(I-Q)/2$. The components 
of the ingoing Stokes parameters have been distinguished by a prime and they depend upon $\mu' = \cos{\vartheta'}$
and $\varphi'$. The Stokes parameters depend upon the (angular) frequency 
$\omega = 2 \pi \nu$. By definition, the scattering matrix connects the outgoing to the ingoing Stokes parameters as:
\begin{equation}
{\mathcal I}^{\mathrm{out}}_{i} (\omega,\,\mu,\,\mu',\,\varphi,\, \varphi') = {\mathcal S}_{i\,j}(\omega,\,\mu,\,\mu',\,\varphi,\, \varphi') 
\, {\mathcal I}^{\mathrm{in}}_{j} (\omega,\,\mu',\, \varphi').
\label{SM}
\end{equation}
The coordinate system has been fixed as\footnote{The orientation of the coordinate system corresponding to Eqs. (\ref{co1})--(\ref{co3}) implies that $\hat{\vartheta} \times \hat{\varphi} = \hat{r}$. In other classic references such as \cite{chandra} 
the orientation is such that $\hat{\vartheta} \times \hat{\varphi} = -\hat{r}$; this different choice 
entails a modification of Eq. (\ref{co3}), i.e.  $\hat{\varphi} = (  \sin{\varphi},\,-\cos{\varphi}, 0)$. The conventions spelled out by Eqs. (\ref{co1})--(\ref{co3}) 
will be followed throughout the paper.}:
\begin{eqnarray}
&& \hat{r} = ( \cos{\varphi} \sin{\vartheta},\, \sin{\varphi} \sin{\vartheta}, \cos{\vartheta}),
\label{co1}\\
&& \hat{\theta} = ( \cos{\varphi} \cos{\vartheta},\, \sin{\varphi} \cos{\vartheta}, -\sin{\vartheta}),
\label{co2}\\
&& \hat{\varphi} = ( - \sin{\varphi},\,\cos{\varphi}, 0).
\label{co3}
\end{eqnarray}
The purpose is to obtain the scattering matrix of electron-photon scattering 
(see, e.g. \cite{chandra}) but in the presence of a magnetic field and in the guiding centre 
approximation \cite{gc1} which is, in practice, a controlled expansion in gradients of the 
magnetic field intensity. 
The derivation of the various components of the scattering matrix is reported in 
appendix \ref{APPA}. In what follows only the results will be reported and discussed. 
Defining as  $r_{\mathrm{e}} = e^2/m_{\mathrm{e}}$ 
the classical radius of the electron, the various components of the scattering matrix can be written as:
\begin{eqnarray}
 {\mathcal S}_{11} &=& \frac{r_{\mathrm{e}}^2}{2 r^2}\biggl\{ 2 \Lambda_{3}(\omega) ( 1 -\mu^2) ( 1 - {\mu'}^2) + 
\zeta^2(\omega) \mu^2\, {\mu'}^2 \biggl[ \Lambda_{1}^2(\omega) + f_{\mathrm{e}}^2(\omega) \Lambda_{2}^2(\omega)\biggr]
\nonumber\\
&-&  4 \zeta(\omega)\Lambda_{1}(\omega) \Lambda_{3}(\omega) \mu \mu' \sqrt{1 - \mu^2} \sqrt{1 - {\mu'}^2} \cos{(\varphi' - \varphi)}
\nonumber\\
&+& 
 \zeta^2(\omega) \biggl[ \Lambda_{1}^2(\omega) - f_{\mathrm{e}}^2(\omega) \Lambda_{2}^2(\omega)\biggr] \mu^2 {\mu'}^2 \cos{[2 (\varphi' - \varphi)]}\biggr\},
\label{P11}\\
 {\mathcal S}_{12}  &=&  \frac{\mu^2 \zeta^2(\omega) r_{\mathrm{e}}^2}{2\,r^2} \biggl\{ \Lambda_{1}^2(\omega) + f_{\mathrm{e}}^2(\omega) \Lambda_{2}^2(\omega) 
- \biggl[\Lambda_{1}^2(\omega) - f_{\mathrm{e}}^2(\omega) \Lambda_{2}^2(\omega)\biggr]\cos{[2(\varphi' - \varphi)]}\biggr\},
\label{P12}\\
 {\mathcal S}_{13}  &=&   \frac{r_{\mathrm{e}}^2}{ r^2} \biggl\{\zeta(\omega) \Lambda_{1}(\omega) \Lambda_{3}(\omega) \mu \sqrt{1 - \mu^2} \sqrt{1 - {\mu'}^2} \sin{(\varphi' - \varphi)} 
\nonumber\\
&-& \frac{1}{2} \zeta^2(\omega) \mu^2 \mu' \biggl[\Lambda_{1}^2(\omega) - f_{\mathrm{e}}^2(\omega) \Lambda_{2}^2(\omega)\biggr] \sin{[2(\varphi'-\varphi)]}\biggr\},
\label{P13}\\
 {\mathcal S}_{14}  &=& \frac{r_{\mathrm{e}}^2f_{\mathrm{e}}(\omega) \zeta(\omega) \Lambda_{2}(\omega)}{r^2} \mu
 \biggl[ \zeta(\omega) \Lambda_{1}(\omega) \mu \mu' - \Lambda_{3}(\omega) 
\sqrt{1 - \mu^2} \sqrt{1 - {\mu'}^2} \cos{(\varphi'-\varphi)}\biggr],
\label{P14}\\
{\mathcal S}_{21}  &=& \frac{\zeta^2(\omega) {\mu'}^2\, r_{\mathrm{e}}^2}{2\,r^2}\biggl\{ \Lambda_{1}^2(\omega) +
f_{\mathrm{e}}^2(\omega) \Lambda_{2}^2(\omega) 
\nonumber\\
&-& \biggl[\Lambda_{1}^2(\omega) - 
f_{\mathrm{e}}^2(\omega) \Lambda_{2}^2(\omega)\biggr] \cos{[2(\varphi'-\varphi)]}\biggr\},
\label{P21}\\
{\mathcal S}_{22}  &=&  \frac{\zeta^2(\omega) \,r_{\mathrm{e}}^2}{2\, r^2}\biggl\{ \Lambda_{1}^2(\omega) 
+ f_{\mathrm{e}}^2(\omega) \Lambda_{2}^2(\omega) 
+\biggl[\Lambda_{1}^2(\omega) - f_{\mathrm{e}}^2(\omega) \Lambda_{2}^2(\omega)\biggr] \cos{[2(\varphi' - \varphi)]}\biggr\},
\label{P22}\\
 {\mathcal S}_{23}  &=&
\frac{\zeta^2(\omega)\, r_{\mathrm{e}}^2}{2\, r^2} \biggl[\Lambda_{1}^2(\omega) - f_{\mathrm{e}}^2(\omega) \Lambda_{2}^2(\omega)
\biggr]\mu' \sin{[2(\varphi' - \varphi)]},
\label{P23}\\
{\mathcal S}_{24}  &=&  \frac{r_{\mathrm{e}}^2}{ r^2}  f_{\mathrm{e}}(\omega) 
\, \zeta^2(\omega) \, \Lambda_{1}(\omega)\, \Lambda_{2}(\omega) \mu',
\label{P24}\\
{\mathcal S}_{31}  &=&\frac{r_{\mathrm{e}}^2}{r^2} \biggl\{- 2 \zeta(\omega) \Lambda_{1}(\omega) \Lambda_{3}(\omega) \sqrt{1 -\mu^2} \mu' \, 
\sqrt{1 - {\mu'}^2} \sin{(\varphi'-\varphi)}
\nonumber\\
&+&\zeta^2(\omega) \mu {\mu'}^2\biggl[ \Lambda_{1}^2(\omega)  
- f_{\mathrm{e}}^2(\omega) \Lambda_{2}^2(\omega) \biggr]\sin{[2(\varphi' - \varphi)]} \biggr\},
\label{P31}\\
{\mathcal S}_{32} &=& - \frac{r_{\mathrm{e}}^2}{r^2}
\zeta^2(\omega) \biggl[ \Lambda_{1}^2(\omega) - f_{\mathrm{e}}^2(\omega) \Lambda_{2}^2(\omega)\biggr] \mu \sin{[2(\varphi' - \varphi)]},
\label{P32}\\
 {\mathcal S}_{33} &=& \frac{ \zeta(\omega) r_{\mathrm{e}}^2}{r^2}  \biggl\{ - \Lambda_1(\omega) 
\Lambda_{3}(\omega) \sqrt{1 - \mu^2} 
\sqrt{1 - {\mu'}^2} \cos{(\varphi' - \varphi)}   
\nonumber\\
&+& \zeta(\omega) \biggl[\Lambda_{1}^2(\omega)- f_{\mathrm{e}}^2(\omega) \Lambda_{2}^2(\omega)\biggr] \mu\, \mu' \cos{[2(\varphi' - \varphi)]}\biggr\},
\label{P33}\\
 {\mathcal S}_{34} &=& - \frac{r_{\mathrm{e}}^2}{r^2} f_{\mathrm{e}}(\omega) \zeta(\omega) \Lambda_{2}(\omega) 
\Lambda_{3}(\omega) \sqrt{1 - \mu^2} \sqrt{1 - {\mu'}^2} \sin{(\varphi' -\varphi)},
\label{P34}\\
 {\mathcal S}_{41} &=& \frac{ r_{\mathrm{e}}^2 f_{\mathrm{e}}(\omega)
\zeta(\omega) \Lambda_{2}(\omega) \mu'}{r^2} \biggl\{ \zeta(\omega) \Lambda_{1}(\omega) \mu \mu'
\nonumber\\
&&- \Lambda_{3}(\omega) \sqrt{1 - \mu^2} \sqrt{1- {\mu'}^2} \cos{(\varphi' - \varphi)} \biggr\},
\label{P41}\\
 {\mathcal S}_{42} &=& \frac{r_{\mathrm{e}}^2 \,f_{\mathrm{e}}(\omega)\zeta^2(\omega) \Lambda_{2}(\omega) \mu }{r^2},
\label{P42}\\
{\mathcal S}_{43} &=& \frac{r_{\mathrm{e}}^2}{r^2}
f_{\mathrm{e}}(\omega) \zeta(\omega) \Lambda_{2}(\omega) \Lambda_{3}(\omega) \sqrt{1 -\mu^2} \sqrt{1 - {\mu'}^2} 
\sin{(\varphi' - \varphi)} 
\label{P43}\\
{\mathcal S}_{44} &=& \frac{r_{\mathrm{e}}^2}{r^2}\biggl\{
\mu \mu' \zeta^2(\omega) \biggl[ \Lambda^2_{1}(\omega)  + f_{\mathrm{e}}^2(\omega)
\Lambda_{2}^2(\omega)\biggr] 
\nonumber\\
&-& \zeta(\omega) \Lambda_{1}(\omega) \Lambda_{3}(\omega) \sqrt{1 - \mu^2} \sqrt{1 - 
{\mu'}^2} \cos{(\varphi' -\varphi)}\biggr\}.
\label{P44}
\end{eqnarray}
In Eqs. (\ref{P11})--(\ref{P44}) various (frequency dependent) 
quantities have been introduced, namely $f_{\mathrm{e}}(\omega)$, $\Lambda_{1}(\omega)$, $\Lambda_{2}(\omega)$ 
and $\Lambda_{3}(\omega)$. Their explicit expressions are:
\begin{eqnarray}
\Lambda_{1}(\omega) &=& 1 + 
\biggl(\frac{\omega^2_{\mathrm{p\,\,i}}}{\omega^2_{\mathrm{p\,\,e}}}\biggr) \biggl( 
\frac{ \omega^2 - \omega^2_{\mathrm{B\,\,e}}}{\omega^2 - \omega^2_{\mathrm{B\,\,i}}}\biggr),
\label{LAM1}\\
\Lambda_{2}(\omega) &=& 1 - \biggl(\frac{\omega^2_{\mathrm{p\,\,i}}}{\omega^2_{\mathrm{p\,\,e}}}\biggr)
\biggl(\frac{\omega_{\mathrm{B\,\,i}}}{\omega_{\mathrm{B\,\,e}}}\biggr) \biggl( 
\frac{ \omega^2 - \omega_{\mathrm{B\,\,e}}^2}{\omega^2 - \omega^2_{\mathrm{B\,\,i}}}\biggr),
\label{LAM2}\\
\Lambda_{3}(\omega) &=& 1 + \biggl(\frac{\omega^2_{\mathrm{p\,\,i}}}{\omega^2_{\mathrm{p\,\,e}}}\biggr),
\label{LAM3}\\
\zeta(\omega) &=& \frac{1}{f_{\mathrm{e}}^2(\omega) -1} = \frac{\omega^2}{\omega_{\mathrm{Be}}^2 - \omega^2}, \qquad 
f_{\mathrm{e}}(\omega) =  \biggl(\frac{\omega_{\mathrm{B\,\,e}}}{\omega}\biggr),
\label{zetaomega}
\end{eqnarray}
where $\omega_{\mathrm{B\,\,e,\,i}}$ and $\omega_{\mathrm{p\,\,e,\,i}}$ are the Larmor and plasma 
frequencies for electrons and ions, namely:
\begin{eqnarray}
\omega_{\mathrm{B\,\,e}} &=& \frac{e\, B}{m_{\mathrm{e}}a}, \qquad 
\omega_{\mathrm{B\,\,i}} = \frac{e\, B}{m_{\mathrm{p}}a},
\label{LARM}\\
\omega_{\mathrm{p\,\,e}} &=& \sqrt{\frac{4 \pi \,e^2\, n_{0}}{m_{\mathrm{e}} \,a(\tau)}}, \qquad 
\omega_{\mathrm{p\,\,i}} = \sqrt{\frac{4 \pi \,e^2\, n_{0}}{m_{\mathrm{p}}\,a(\tau)}},
\label{PLAS}
\end{eqnarray}
where $B$ is the modulus of the magnetic field intensity coinciding, in practice, with the lowest order result of the guiding centre approximation.
As discussed in appendix \ref{APPA},
Eqs. (\ref{LARM}) and (\ref{PLAS}) take into account the redshift dependence of the frequency.
The magnetic field appearing in Eq. (\ref{LARM}) is the comoving magnetic field (see appendix \ref{APPA})
and this means, in particular,  that the relation to the physical frequencies is given by
\begin{equation}
\omega_{\mathrm{B\,\,e}} = \frac{ e {\mathcal B}}{m_{\mathrm{e}}} a(\tau) = 
\omega_{\mathrm{B\,\,e}}^{\mathrm{phys}} a(\tau),
\label{PLAS2}
\end{equation}
and similarly for the other quantities of Eqs. (\ref{LARM}) and (\ref{PLAS}).
The logic behind the functions defined in Eqs. (\ref{LAM1}), (\ref{LAM2}) and (\ref{LAM3}) is that 
we want to factorize the electron contribution by keeping track of the ions. According to this strategy, to leading order in the ion contributions, $\Lambda_{1}(\omega)$, $\Lambda_{2}(\omega)$ and 
$\Lambda_{3}(\omega)$ turn out to be:
\begin{eqnarray} 
\Lambda_{1}(\omega) &=&  1 + {\mathcal O}\biggl(\frac{m_{\mathrm{e}}}{m_{\mathrm{p}}}\biggr) \bigg[ 1 + {\mathcal O}\biggl(\frac{\omega_{\mathrm{B\,\,e}}}{\omega}\biggr)\biggr],
\nonumber\\
\Lambda_{2}(\omega) &=&1 - {\mathcal O}\biggl(\frac{m^2_{\mathrm{e}}}{m^2_{\mathrm{p}}}\biggr) \bigg[ 1 + {\mathcal O}\biggl(\frac{\omega_{\mathrm{B\,\,e}}}{\omega}\biggr)\biggr],
\nonumber\\
\Lambda_{2}(\omega) &=& 1 +  {\mathcal O}\biggl(\frac{m_{\mathrm{e}}}{m_{\mathrm{p}}}\biggr).
\label{LAMexp}
\end{eqnarray}
In the limit $f_{\mathrm{e}}(\omega) \to 0$ and by correspondingly setting $\Lambda_{1}(\omega) = \Lambda_{2}(\omega) = \Lambda_{3}(\omega) =1$,  Eqs. (\ref{P11})--(\ref{P44}) reproduce exactly 
the results of Ref. \cite{chandra} modulo the different orientation 
of the coordinate system (see Eqs. (\ref{co1})--(\ref{co3}) and footnote therein) which leads to an overall sign difference 
in the matrix elements containing the sines. In the limit $\Lambda_{1}(\omega) = \Lambda_{2}(\omega) = \Lambda_{3}(\omega) \to 1$ the present 
results coincide  with Ref. \cite{TS1} (see also \cite{TS2} modulo the typo pointed out in \cite{TS1}). 
The evolution equations for the brightness perturbations can be written, in general terms, as
\footnote{Note the presence of the Doppler term  arising in the source term
of the intensity. The collisionless part of the evolution of the intensity perturbations is well known. For 
the interested reader it is derived, within the present conventions, in Ref. \cite{primer} and within different 
gauge choices.}
\begin{eqnarray}
&& \Delta_{\mathrm{I}}' + \epsilon' \Delta_{\mathrm{I}} + n^{i} \partial_{i} \Delta_{\mathrm{I}} = 
\psi' - n^{i} \partial_{i} \phi + \epsilon' \mu v_{b}+ \epsilon' C_{\mathrm{I}}(\mu,\omega),
\label{Br1}\\
&&\Delta_{\mathrm{Q}}' + \epsilon' \Delta_{\mathrm{Q}} + n^{i} \partial_{i} \Delta_{\mathrm{Q}} = \epsilon' C_{\mathrm{Q}}(\mu,\omega),
\label{Br2}\\
&& \Delta_{\mathrm{U}}' + \epsilon' \Delta_{\mathrm{U}} + n^{i} \partial_{i} \Delta_{\mathrm{U}} = \epsilon' C_{\mathrm{U}}(\mu,\omega),
\label{Br3}\\
&& \Delta_{\mathrm{V}}' + \epsilon' \Delta_{\mathrm{V}} + n^{i} \partial_{i} \Delta_{\mathrm{V}}  = \epsilon' C_{\mathrm{V}}(\mu,\omega).
\label{Br4}
\end{eqnarray}
Concerning Eqs. (\ref{Br1})--(\ref{Br4}) few comments are in order.  The brightness perturbations can be classified in terms 
of their transformation properties under rotations in the three-dimensional Euclidean sub-manifold. This 
means that all the brightness perturbations will have a scalar, a vector and a tensor contribution. In the framework of 
the $\Lambda$CDM scenario the tensor and the vector fluctuations of the geometry are totally absent. Therefore, since 
we want to compute the circular polarization in the minimal situation, we will stick to the scalar modes 
of the geometry which are connected with the scalar modes of the brightness perturbations. As implied by the concordance model the 
background metric is taken to be conformally flat with signature 
mostly minus, i.e. $\overline{g}_{\mu\nu} = a^2(\tau) \eta_{\mu\nu}$ where 
$\eta_{\mu\nu} = \mathrm{diag}(1,\,-1,\, -1,\, -1)$ is the Minkowski metric.
In Eq. (\ref{Br1}),  $\phi$ and $\psi$ 
represent the scalar  fluctuations of the metric in the conformally Newtonian gauge, i.e. 
$\delta_{\mathrm{s}} g_{00} = 2 a^2 \phi$ and $\delta_{\mathrm{s}} g_{ij} = 2 a^2 \delta_{ij} \psi$. Always 
 Eq. (\ref{Br1}), $v_{\mathrm{b}}$ is related to the baryon velocity (see appendix \ref{APPA}) i.e. 
 the center of mass velocity of the electron-ion system. In Eqs. (\ref{Br1})--(\ref{Br4}) 
  $\epsilon'$ denotes, as usual, the differential optical depth, i.e. 
\begin{equation}
\epsilon' = x_{\mathrm{e}} \,\tilde{n}_{\mathrm{e}}\, \sigma_{\gamma\mathrm{e}} \frac{a}{a_{0}}, \qquad 
\sigma_{\gamma\mathrm{e}} = \frac{8}{3} \pi r_{\mathrm{e}}^2.
\label{Br5}
\end{equation}
After integration over $\varphi'$ the source terms appearing in 
Eqs. (\ref{Br1})--(\ref{Br4}) can be written as 
\begin{eqnarray}
C_{\mathrm{I}}(\mu,\omega) &=&  \frac{3}{16\pi}\int_{-1}^{1} d\mu' 
\biggl\{ \biggl[\overline{P}_{11} + \overline{P}_{12} + \overline{P}_{21} + \overline{P}_{22}\biggr] \Delta_{\mathrm{I}}(\mu,\mu') + 
\nonumber\\
&+& \biggl[\overline{P}_{11} - \overline{P}_{12} + \overline{P}_{21} - \overline{P}_{22}\biggr] 
\Delta_{\mathrm{Q}}(\mu,\mu')  
+ 2 \biggl[\overline{P}_{13} + \overline{P}_{23}\biggr] \Delta_{\mathrm{U}}(\mu,\mu') 
\nonumber\\
&+&  2 \biggl[\overline{P}_{14} + \overline{P}_{24}\biggr] \Delta_{V}(\mu,\mu') \biggr\},
\label{CI}\\
C_{\mathrm{Q}}(\mu,\omega) &=& \frac{3}{16 \pi}\int_{-1}^{1} d\mu' 
\biggl\{ \biggl[\overline{P}_{11} + \overline{P}_{12} - \overline{P}_{21} - \overline{P}_{22}\biggr] \Delta_{\mathrm{I}}(\mu,\mu')+ 
\nonumber\\
&+& \biggl[\overline{P}_{11} - \overline{P}_{12} - \overline{P}_{21} + \overline{P}_{22}\biggr] \Delta_{\mathrm{Q}}(\mu,\mu') 
\nonumber\\
&+& 2 \biggl[\overline{P}_{13} - \overline{P}_{23}\biggr] \Delta_{\mathrm{U}} + 2 \biggl[\overline{P}_{14} - 
\overline{P}_{24}\biggr] \Delta_{\mathrm{V}}(\mu,\mu')\biggr\},
\label{CQ}\\
C_{\mathrm{U}}(\mu,\omega) &=& \frac{3}{16 \pi}\int_{-1}^{1} d\mu' \biggl\{\biggl[\overline{P}_{31} + \overline{P}_{32}\biggr]
\Delta_{\mathrm{I}}(\mu,\mu') 
+ \biggl[\overline{P}_{31} - \overline{P}_{32}\biggr]\Delta_{\mathrm{Q}}(\mu,\mu') 
\nonumber\\
&+&2 \overline{P}_{33}\Delta_{\mathrm{U}}(\mu,\mu') + 2 \overline{P}_{34} \Delta_{\mathrm{v}}(\mu,\mu')\biggr\},
\label{CU}\\
C_{\mathrm{V}}(\mu,\omega) &=& \frac{3}{16 \pi}\int_{-1}^{1} d\mu' \biggl\{\biggl[\overline{P}_{41} + \overline{P}_{42}\biggr]
\Delta_{\mathrm{I}}(\mu,\mu') 
+ \biggl[\overline{P}_{41} - \overline{P}_{42}\biggr]\Delta_{\mathrm{Q}}(\mu,\mu') 
\nonumber\\
&+& \overline{P}_{43}\Delta_{\mathrm{U}}(\mu,\mu') + \overline{P}_{44} \Delta_{\mathrm{V}}(\mu,\mu')\biggr\},
\label{CV}
\end{eqnarray}
where the generic matrix element appearing in Eqs. (\ref{CI})--(\ref{CV}) is the integral  over $\varphi'$ 
of the corresponding element of the scattering matrix of Eq. (\ref{SM}), i.e. 
\begin{equation}
\overline{P}_{ij}(\mu,\mu') = \int_{0}^{2 \pi} \frac{r^2}{r_{\mathrm{e}}^2}\, {\mathcal S}_{ij}(\mu, \mu', \varphi') \,  \,d\varphi'.
\end{equation}
For immediate convenience it is appropriate to write down the explicit form 
of the various matrix elements appearing in Eqs. (\ref{CI})--(\ref{CV}):
\begin{eqnarray}
\overline{P}_{11}(\omega,\mu,\mu') &=& \pi\biggl\{ 2 \Lambda_{3}^2(\omega) (1 - \mu^2) (1- {\mu'}^2) + 
\zeta^2(\omega) \mu^2 \, {\mu'}^2 \biggl[\Lambda_{1}^2(\omega) + f_{\mathrm{e}}^2(\omega) \Lambda_{2}^2(\omega) \biggl] \biggr\},
\nonumber\\
\overline{P}_{12}(\omega,\mu,\mu') &=& \pi \mu^2 \zeta^2(\omega) \biggl[  \Lambda_{1}^2(\omega) + f_{\mathrm{e}}^2(\omega) \Lambda_{2}^2(\omega)\biggr],
\nonumber\\
\overline{P}_{13}(\omega,\mu,\mu') &=& 0,
\nonumber\\
\overline{P}_{14}(\omega, \mu,\mu') &=& 2\pi f_{\mathrm{e}}(\omega)
\zeta^2(\omega) \Lambda_{1}(\omega) \Lambda_{2}(\omega) \mu^2 \mu',
\nonumber\\
\overline{P}_{21}(\omega,\mu,\mu') &=& \pi \zeta^2(\omega) {\mu'}^2 [ \Lambda_{1}^2(\omega) + 
f_{\mathrm{e}}^2(\omega) \Lambda_{2}^2(\omega)],
\nonumber\\
\overline{P}_{22}(\omega, \mu,\mu') &=& \pi\zeta^2 [\Lambda_{1}^2(\omega) + f_{\mathrm{e}}^2(\omega) \Lambda_{2}^2(\omega)],
\nonumber\\
\overline{P}_{23}(\omega,\mu,\mu') &=& 0, 
\nonumber\\
\overline{P}_{24}(\omega,\mu,\mu') &=& 2\pi  f_{\mathrm{e}}(\omega) \zeta^2(\omega) \Lambda_{1}(\omega)
 \Lambda_{2}(\omega) \mu',
\nonumber\\
\overline{P}_{31}(\omega,\mu,\mu') &=& \overline{P}_{32}(\omega,\mu,\mu')=\overline{P}_{33}(\omega,\mu,\mu')=
\overline{P}_{34}(\omega,\mu,\mu')=0,
\nonumber\\
\overline{P}_{41}(\omega,\mu,\mu') &=& 2\pi f_{\mathrm{e}}(\omega) \zeta^2(\omega) \Lambda_{1}(\omega) \Lambda_{2}(\omega) \mu {\mu'}^2,
\nonumber\\
\overline{P}_{42}(\omega,\mu,\mu') &=& 2\pi f_{\mathrm{e}}(\omega) \zeta^2(\omega) \Lambda_{2}(\omega) \Lambda_{1}(\omega) \mu ,
\nonumber\\
\overline{P}_{43}(\omega,\mu,\mu') &=&0,
\nonumber\\
\overline{P}_{44}(\omega,\mu,\mu') &=& 2\pi \zeta^2(\omega) \biggl[\Lambda^2_{1}(\omega) + 
f_{\mathrm{e}}^2(\omega)  \Lambda_{2}^2(\omega) \biggr] \mu \mu'.
\label{INTS}
\end{eqnarray}
Inserting the results of Eq. (\ref{INTS}) inside Eqs. (\ref{CI})--(\ref{CV})
the explicit expressions of the various source terms can be obtained. 
The details of this standard manipulation are reported in appendix \ref{APPB}
(see, in particular, Eqs. (\ref{interm1})--(\ref{integral1})). The final result is:
\begin{eqnarray}
C_{\mathrm{I}}(\omega,\mu) &=&
 \frac{1}{4}\biggl\{\Delta_{\mathrm{I}0} \biggl[ 2 \Lambda_{3}(\omega) (1 - \mu^2) + 2 \zeta^2(\omega) \biggl(\Lambda_{1}^2(\omega) + f_{\mathrm{e}}^2 (\omega) \Lambda_{2}^2(\omega) \biggr)(1 +\mu^2) \biggr]
 \nonumber\\
 &+& \biggl[ 2 \Lambda_{3}(\omega) (1 - \mu^2) - \zeta^2(\omega)  \biggl(\Lambda_{1}^2(\omega) + f_{\mathrm{e}}^2 (\omega) \Lambda_{2}^2(\omega) \biggr)( 1 + \mu^2) \biggr] S_{\mathrm{P}}
 \nonumber\\
 &-& 6\, i \,f_{\mathrm{e}}(\omega) \zeta^2(\omega) \Lambda_{1}(\omega) \Lambda_{2}(\omega)(1 + \mu^2)\Delta_{\mathrm{V}1}\biggr\},
 \label{CIF}\\
C_{\mathrm{Q}}(\omega,\mu) &=& \frac{1}{4}\biggl\{ \Delta_{\mathrm{I}0} 
\biggl[ 2 (1 - \mu^2)\Lambda_{3}(\omega) - 2\zeta^2(\omega)  \biggl(\Lambda_{1}^2(\omega) + f_{\mathrm{e}}^2 (\omega) \Lambda_{2}^2(\omega) \biggr)( 1 - \mu^2)   \biggr]
 \nonumber\\
 &+& \biggl[ 2 \Lambda_{3}(\omega) (1 - \mu^2) + \zeta^2(\omega) \biggl(\Lambda_{1}^2(\omega) + f_{\mathrm{e}}^2 (\omega) \Lambda_{2}^2(\omega) \biggr)( 1 - \mu^2) \biggr] S_{\mathrm{P}}
 \nonumber\\
&-& 6 i f_{\mathrm{e}}(\omega) \zeta^2(\omega) \Lambda_{2}(\omega)  \Lambda_{1}(\omega)(\mu^2 -1)\Delta_{\mathrm{V}1}\biggr\},
 \label{CQF}\\
C_{\mathrm{V}}(\omega,\mu) &=& \frac{\zeta^2(\omega) P_{1}(\mu)}{2}\biggl\{f_{\mathrm{e}}(\omega) \Lambda_{2}(\omega) 
\Lambda_{1}(\omega)\biggl[ 2 \Delta_{\mathrm{I}0} - S_{\mathrm{P}} \biggr] 
  \nonumber\\
  &-& \frac{3}{2} i \biggl[ \Lambda_{1}^2(\omega) + f_{\mathrm{e}}^2(\omega) \Lambda_{2}^2(\omega)\biggr] \Delta_{\mathrm{V}1}\biggr\},
\label{CVF}
 \end{eqnarray}
 while $C_{\mathrm{U}}(\omega,\mu)$ vanishes identically. In Eqs. (\ref{CIF}), (\ref{CQF}) and (\ref{CVF}) 
 the following important combination has been introduced, namely:
\begin{equation}
S_{\mathrm{P}} = \Delta_{\mathrm{I}2} +  \Delta_{\mathrm{Q}2} + \Delta_{\mathrm{Q}0}.
\label{SP}
\end{equation}
It must be remarked that $S_{\mathrm{P}}$ is the standard source term arising in the treatment of CMB polarization when the 
magnetic field contribution is ignored in the scattering process (see, e.g. \cite{max1,sp1,sp2} and references therein).
The result expressed by Eqs. (\ref{CIF}), (\ref{CQF}) and (\ref{CVF})  does 
hold to lowest order in the guiding centre approximation (see  Eq. (\ref{co6a})
and discussion therein). As it will be discussed in \ref{sec4}, 
the results derived so far improve the accuracy of the radiative transfer
equations in the case when large-scale magnetic fields are 
consistently included in the discussion. 

The Faraday effect of the CMB is just a rotation of the linear polarization 
of the CMB and it does not involve the generation of any circular polarization. Faraday effect has been 
recently treated in greater detail by including various effects which have been neglected in the past \cite{max3} (see also 
\cite{max5}). In the case of the Faraday effect, first the linear polarization is generated because
of the quadrupole in the intensity of the radiation field and then the polarization is rotated. Also Faraday 
rotation is treated often in the uniform field approximation but without the explicit contribution 
of the magnetic field intensity to the scattering. The present formulation
improves also on the treatment of Faraday effect of the CMB (see \cite{rev2} for an introduction)  even if, to keep the discussion 
self-contained, the focus will be on the generation of the V-mode polarization. 

In connection with the Faraday rotation it is appropriate to mention the different physical nature 
of the approximations often employed in the discussion of large-scale magnetism. In the present paper, as 
already mentioned, the guiding centre approximation has been employed. This approximation (see appendix \ref{APPA}) 
is particularly sound in the case of scattering problems when the wavelength of the scattered photons is 
much shorter than the inhomogeneity scale of the magnetic field \cite{gc1}.  The guiding centre approximation 
does not break explicitly the isotropy of the background. As already mentioned in the previous paragraph, 
Faraday rotation can be discussed in the uniform field approximation \cite{far1,far2,far3,far4}. The uniform 
field approximation is independent upon the guiding centre approximation: indeed, for instance, 
in the studies of \cite{far1,far2,far3,far4} the magnetic field does not contribute to the scattering 
matrix while it does rotate the polarization plane of the CMB. The uniform field approximation 
holds provided the magnetic field is not too strong. In the latter case a (new) preferred direction in the sky 
pops up; a potential correlation of the $a^{\mathrm{T}}_{\ell -1,m}$ and 
$a^{\mathrm{T}}_{\ell +1,m}$ multipole coefficients is induced and, from this 
observation, uniform magnetic fields can be constrained \cite{unif1,unif2,unif3,unif4}. This last case 
is not directly related to the present considerations. 

\renewcommand{\theequation}{3.\arabic{equation}}
\setcounter{equation}{0}
\section{Circular polarization from magnetic knots}
\label{sec3}
It is appropriate to highlight a possible connection between the occurrence 
of circular polarization and the topological properties of the magnetic flux lines. The occurrence 
of circular polarization is directly related  to the Lorentz force 
acting either on the individual charge carriers  (i.e. $\vec{v}_{\mathrm{e,\, i}} \times \vec{B}$) or on the Ohmic current (i.e. $\vec{J} \times \vec{B}$).  In a plasma characterized by a finite conductivity, the presence (or absence) of Lorentz force term can be directly related to  the topology of the magnetic flux lines.
The topology of the magnetic flux lines can be classified in terms of the so-called magnetic helicity, i.e. 
\begin{equation}
{\mathcal N}_{B} = \int_{V} d^{3} x \,\vec{A} \cdot \vec{B}, 
\label{hel1}
\end{equation}
where $\vec{A}$ is the vector potential; Eq. (\ref{hel1}) is the magnetohydrodynamical analog of the kinetic helicity, i.e. 
\begin{equation}
{\mathcal N}_{v} = \int d^{3} x \,\vec{v} \cdot \vec{\omega}, \qquad \vec{\omega} = \vec{\nabla} \times \vec{v}
\label{hel2}
\end{equation}
where $\vec{\omega}$ is the vorticity.
In simply connected domains, the magnetic helicity is gauge-invariant provided the normal component 
of $\vec{B}$ vanishes at the boundary surface of the integration volume $V$. Furthermore, the 
helicity is also gauge-invariant if the integration volume is given by a single (or multiple) magnetic flux tube. 
An important property is that the magnetic helicity is conserved in a highly conducting plasma.
In particular, it can be shown, that (see, e.g. \cite{rev1} or the seminal paper of \cite{wolt})
\begin{equation}
\frac{d {\mathcal N}_{B}}{d \tau} = - \frac{1}{4 \pi \sigma} \int_{V} d^{3} x \vec{B} \cdot \vec{\nabla} \times \vec{B}, 
\label{hel3}
\end{equation}
where $\sigma$ is the conductivity. In the limit $\sigma\to \infty$ the magnetic helicity is exactly 
conserved. 
In minimizing the total magnetic energy with the constraint that the magnetic helicity be conserved we are naturally 
led to the variational problem 
\begin{equation}
\delta \biggl[ \int_{V} d^{3} x \biggl(|\vec{B}|^2  - \alpha \vec{A} \cdot \vec{B} \biggr)\biggr]=0,
\label{hel4}
\end{equation}
where $\alpha$ is a Lagrange multiplier. Since $\vec{B} = \vec{\nabla} \times \vec{A}$, the variational problem of Eq. (\ref{hel4}) is equivalent to 
\begin{equation}
\delta \biggl[ \int_{V} d^{3} x \biggl( |\vec{\nabla}\times \vec{A}|^2  - \alpha \vec{A} \cdot \vec{\nabla}\times \vec{A} \biggr)\biggr]=0.
\label{hel5}
\end{equation}
By making the variation explicit, we have that Eq. (\ref{hel5}) implies that 
\begin{equation}
\vec{\nabla}\times \vec{\nabla}\times \vec{A} = \alpha \vec{\nabla} \times \vec{A}.
\label{hel6}
\end{equation}
Going back to the magnetic field we have that the configurations 
\begin{equation}
\vec{\nabla}\times \vec{B} = \alpha \vec{B}
\label{hel7}
\end{equation}
correspond to the lowest state of magnetic field energy which a closed system may attain. The variational 
approach leading to the condition (\ref{hel7}) is due to Woltjer \cite{wolt} (see also, in this connection \cite{fermi1,fermi2} and \cite{ch1,ch2}). The configurations obeying Eq. (\ref{hel7}) are closely related to the concept of magnetic knot \cite{kn1}
and may arise a consequence of the dynamics of the electroweak phase transition. 
The presence of pseudo-scalar interactions at the electroweak time 
can twist the magnetic flux lines of the hypermagnetic field and produce 
a primordial background of hypermagnetic knots \cite{kn2,kn3}. It has been speculated 
that baryogenesis can be related to the presence of hypermagnetic knots and a 
similar way of thinking has been pursued in \cite{vach1,gar1}.
The themes discussed in \cite{kn1,kn2} stimulated 
various investigations both on the dynamics of particles in hypermagnetic knot
configurations \cite{hk1,hk2,hk3} as well as on related ideas \cite{hk4,hk5}. 

In solar spectropolarimetry magnetic knots appear since the late sixties \cite{sk1} (see also, for instance, \cite{sk2}).
An example of magnetic knot configuration is given by:
\begin{equation}
\vec{B}(z) = B_{0}(\sin{k z} \hat{x} + \cos{k z} \hat{y}),
\label{kn1}
\end{equation}
satisfying the force free condition discussed before and  sometimes 
analyzed in connection with the polarization properties of the synchrotron emission \cite{se1,se2}.
The idea is to distinguish the topology of the magnetic flux lines from the 
analysis of the circular polarization. In other words: measurements of circular 
polarization can be used to infer not only the potential existence 
of large-scale magnetic fields but also their topological structure.
As before we are considering here the situation where the ingoing Stokes 
parameters have no azimuthal dependence. The problem will now be to compute 
the various entries of the tensor $P_{ij}(\mu,\mu',z)$ which  
also depend upon the  inhomogeneity scale of the knot.

Consider, for simplicity, the case when protons are neglected and only the leading terms 
are kept in $f_{\mathrm{e}}(\omega)$. In this case we have, quite simply, that 
the phase matrix, after integration over $\varphi'$, will be
\begin{eqnarray}
\overline{P}_{11}(\mu,\mu',z) &=& \pi[ 2 (1 - \mu^2) (1- {\mu'}^2) + 
 \mu^2 \, {\mu'}^2 ] 
\nonumber\\
&+& \pi f_{\mathrm{e}}(\omega)\, [ 3(1 - \mu^2) (1- {\mu'}^2)  +1]  \cos{[2(k z + \varphi)]},
\nonumber\\
\overline{P}_{12}(\mu,\mu',z) &=& \pi \mu^2  + \pi \{1 +  \mu^2 \cos{[2(k z + \varphi)]}\} f_{\mathrm{e}}^2(\omega),
\nonumber\\
\overline{P}_{13}(\mu,\mu',z) &=& 0,
\nonumber\\
\overline{P}_{14}(\mu,\mu', z) &=& 2 \pi \mu  \mu' \sqrt{1 - \mu^2} f_{\mathrm{e}}(\omega) \sin{[k z + \varphi]},
\nonumber\\
\overline{P}_{21}(\mu,\mu',z) &=& \pi {\mu'}^2 + \pi \{1 -  \cos{[2(kz + \varphi)]}\}f_{\mathrm{e}}^2(\omega) ,
\nonumber\\
\overline{P}_{22}(\mu,\mu',z) &=& \pi + \pi \{ 1 - \cos{[2(k z - \varphi)]} \}f_{\mathrm{e}}^2(\omega),
\nonumber\\
\overline{P}_{23}(\mu,\mu',z) &=& 0, 
\nonumber\\
\overline{P}_{24}(\mu,\mu',z) &=& 0
\nonumber\\
\overline{P}_{31}(\mu,\mu',z) &=& - 2 \pi \mu f_{\mathrm{e}}^2(\omega) \sin{[2(k z +\varphi)]},
\nonumber\\
\overline{P}_{32}(\mu,\mu',z) &=& 0,
\nonumber\\
\overline{P}_{33}(\mu,\mu',z) &=& 0,
\nonumber\\
\overline{P}_{34}(\mu,\mu',z) &=& 2\pi \sqrt{1 - \mu^2} \mu' \cos{[k z + \varphi]} f_{\mathrm{e}}(\omega),
\nonumber\\
\overline{P}_{41}(\mu,\mu',z ) &=&  - 2\pi \sqrt{1 - \mu^2} ({\mu'}^2 -2) f_{\mathrm{e}}(\omega) \sin{[k z+ \varphi]},
\nonumber\\
\overline{P}_{42}(\mu,\mu',z) &=& 2 \pi \sqrt{1 - \mu^2}  f_{\mathrm{e}}(\omega) \sin{[k z+ \varphi]},
\nonumber\\
\overline{P}_{43}(\mu,\mu',z) &=&0
\nonumber\\
\overline{P}_{44}(\mu,\mu',z) &=& 2 \pi \mu \mu' [1+ f_{\mathrm{e}}^2(\omega)] .
\nonumber
\end{eqnarray}
By integrating over $z$ the relevant matrix elements, i.e. 
\begin{equation}
\int_{0}^{2\pi/k} \overline{P}_{41}(\mu,\mu', z) dz =0,\qquad \int_{0}^{2\pi/k} \overline{P}_{42}(\mu,\mu', z) dz =0, \qquad \int_{0}^{2\pi/k} \overline{P}_{43}(\mu,\mu', z) dz =0. 
\end{equation}
The same integration, but applied to the linear polarizations, leads to a non-vanishing 
result. Indeed, the coupling of $\Delta_{I}$ to $\Delta_{Q}$ is controlled 
by the following matrix element:
\begin{eqnarray}
&& \overline{P}_{11} + \overline{P}_{12} - \overline{P}_{21} - \overline{P}_{22} = 
\pi( 1 - \mu^2)- 3 \pi{\mu'}^2 (1 - \mu^2)
\nonumber\\
&& + \pi \{ 3 ( 1 - \mu^2) (1 - {\mu'}^2) + 2 ( 1+ \mu^2) \cos{[2(k z + \varphi)]}\} f_{\mathrm{e}}^2 
\end{eqnarray}
whose integral over $z$ does not vanish and is given by
\begin{equation}
\frac{\pi^2}{k}( 1 - \mu^2) ( 2 - 6 {\mu'}^2) +\frac{2 \pi^2}{k} (1 -\mu^2)[ 1 + 3 {\mu'}^2 + 2 ( 1 - 3 {\mu'}^2)] f_{\mathrm{e}}^2
\end{equation}   
The rationale for this occurrence stems from the fact that magnetic knots 
minimize the magnetic energy subject to the constraint the the helicity 
is constant.  Indeed, over large scales, the minimization 
of the magnetic energy subjected to the constraint that the magnetic helicity 
is conserved is equivalent to the condition 
\begin{equation}
\vec{\nabla}\times \vec{B} = \frac{\vec{B}}{L}
\label{FIRST}
\end{equation}
where $L$ is has dimensions of a length and denotes the typical scale of the knot.  
Over very large-scales the displacement current can be neglected and, therefore, 
\begin{equation}
\vec{J} = \frac{1}{4\pi} \vec{\nabla} \times \vec{B}, \qquad \vec{J} = e n_{0} (\vec{v}_{\mathrm{i}} - \vec{v}_{\mathrm{e}})
\label{SECOND}
\end{equation}
But because of Eq. (\ref{FIRST}) $\vec{J} = \vec{B}/(4\pi L)$ and, therefore, $\vec{J} \times \vec{B} =0$. This shows that, rather generically, the vanishing of the Lorentz force implies, for these configurations, the vanishing of the circular polarization. 

There is another (indirect) way of appreciating this point. It is well known 
that, at finite conductivity and finite electron density it is possible to construct 
solutions of the Maxwell equations  whose Poynting vector exactly vanishes 
both in the low frequency and in the high-frequency limit, i.e. $\vec{E} \times \vec{B} =0$. These 
solutions are often dubbed helicity waves since they do not carry momentum 
but rather helicity \cite{hw1,hw2,hw3} (see also \cite{hw3a,hw3b}).  Consider first helicity waves in vacuo. 
A consistent solution of Maxwell's equations can be written, in this case, as:
\begin{eqnarray}
\vec{E}(z, \tau) &=& B_{0}\frac{k}{\omega} [ \sin{(k z)} \sin{(\omega \tau)} \hat{x} + \cos{(k z)} \cos{(\omega \tau)} \hat{y}],
\nonumber\\
 \vec{B}(z, \tau) &=& B_{0} [ \sin{(k z)} \cos{(\omega \tau)} \hat{x} + \cos{(k z)} \cos{(\omega \tau)} \hat{y}].
\nonumber
\end{eqnarray}
The above example can also be written as the superposition of circularly polarized waves 
propagating in opposite directions:
\begin{equation}
\vec{E}(z,\tau) = \frac{k \,B_{0}}{2\omega} \{ [\cos{( k z - \omega \tau)} - \cos{(k z + \omega \tau)}] \hat{x} + 
[\sin{(k z + \omega \tau)} - \sin{( k z - \omega \tau)}]\hat{y}\}.
\nonumber
\end{equation}
This class of solution can also be obtained at finite electron density and the pertinent dispersion relations are 
\begin{equation}
\frac{k^2}{\omega^2} = 1 - \frac{\omega_{\mathrm{pe}}^2}{\omega (\omega + i \Gamma)}
\end{equation}
where $\Gamma$ is the collision frequency (i.e. the rate of interactions).
The Ohm law can be written as:
\begin{equation}
\vec{J}= \frac{e^2 n_{\mathrm{e}}}{m_{\mathrm{e}}}\frac{\vec{E}}{(\Gamma - i \omega)}\simeq \sigma \vec{E}, \qquad 
\sigma = \frac{\omega_{\mathrm{pe}}^2}{4 \pi \Gamma}.
\end{equation}
In the limit $\omega \to 0$ the electric fields are suppressed by the conductivity while the magnetic fields will tend 
towards the force-free configuration recalled before. 

The configuration discussed here has some 
realistic features and the most relevant drawback is that it  is not localized in space. Localized knot configurations 
can however be constructed (see \cite{kn3} and references therein and also \cite{JP}). It will be interesting to understand 
the scattering of photons also in these more realistic cases. In spite of that the physical message of the present exercise 
seems to be that circular polarization of the outgoing radiation is generic provided
the underlying magnetic field does affect charged particles. As we saw such an inference 
is not automatic as long as knotted configurations maximize helicity but minimize the Lorentz force. In the latter 
case the scattering matrix might not be affected by the magnetic field if the correlation scale of the magnetic knot is much shorter than the Hubble radius at recombination. 

\renewcommand{\theequation}{4.\arabic{equation}}
\setcounter{equation}{0}
\section{Estimates of the circular polarization}
\label{sec4}
Building up on the results of section \ref{sec3} and taking into account the consideration of section \ref{sec4} it is 
now appropriate to solve the evolution equations of the brightness perturbations and to obtain explicit estimates of the V-mode power spectra.
Since $m_{\mathrm{e}}/m_{\mathrm{p}}\ll 1$ and 
$f_{\mathrm{e}}(\omega) \ll 1$, Eqs. (\ref{CIF}), (\ref{CQF}) and (\ref{CVF}) can be safely expanded 
 in powers of $(m_{\mathrm{e}}/m_{\mathrm{p}})$ as well as in powers of $f_{\mathrm{e}}(\omega)$. The expansion of the scattering matrix in powers of $(m_{\mathrm{e}}/m_{\mathrm{p}})$ is rather common (already in the absence of any 
 magnetic fields) since, to leading order, the mean free  path of the 
photons is chiefly determined by the scattering on the electrons. The expansion in $(m_{\mathrm{e}}/m_{\mathrm{p}})$) is common practice in Boltzmann solvers (see, e.g. \cite{BER}). In the 
present case the same strategy will be employed by adding, however, a further expansion parameter, i.e. $f_{\mathrm{e}}(\omega)$.

While the evolution equations of the brightness perturbations for the intensity and for the linear polarization have the first relevant correction going 
as $f_{\mathrm{e}}^2(\omega)$, the evolution equation  for $\Delta_{\mathrm{V}}$ has a source term proportional to $f_{\mathrm{e}}(\omega)$.
The three functionals appearing in Eqs. (\ref{CIF}), (\ref{CQF}) and (\ref{CVF})  can then be expanded 
in powers of $(m_{\mathrm{e}}/m_{\mathrm{p}})$ and $f_{\mathrm{e}}(\omega)$ with the result that
\begin{eqnarray}
C_{\mathrm{I}}(\omega,\mu) &=& \Delta_{\mathrm{I}0}
\biggl\{ 1 + \biggl[P_{2}(\mu) +2\biggr] f_{\mathrm{e}}^2(\omega)\biggr\} -\frac{S_{\mathrm{P}}}{2}\biggl\{ P_{2}(\mu)  + \biggl[ 2 + P_{2}(\mu)\biggr] f_{\mathrm{e}}^2(\omega) \biggr\} 
\nonumber\\
&-&  i f_{\mathrm{e}}^2(\omega)\biggl[ 2 + P_{2}(\mu)\biggr] \Delta_{\mathrm{V}1}  + {\mathcal O}\biggl(\frac{m_{\mathrm{e}}}{m_{\mathrm{p}}}\biggr) + {\mathcal O}(f_{\mathrm{e}}^4),
\label{CexI1}\\
C_{\mathrm{Q}}(\omega,\mu) &=& \frac{1 -P_{2}(\mu)}{2} \biggl\{S_{\mathrm{P}} + 
f_{\mathrm{e}}^2(\omega)\biggl[ S_{\mathrm{P}} - 2 \Delta_{\mathrm{I}0}  + 2 i \Delta_{\mathrm{V}1}\biggr]\biggr\} 
\nonumber\\
&+&  {\mathcal O}\biggl(\frac{m_{\mathrm{e}}}{m_{\mathrm{p}}}\biggr) + 
{\mathcal O}(f_{\mathrm{e}}^4),
\label{CexQ1}\\
C_{\mathrm{V}}(\omega,\mu) &=& \frac{P_{1}(\mu)}{2} \biggl\{ 2 f_{\mathrm{e}}(\omega) 
[ 2 \Delta_{\mathrm{I}0} - S_{\mathrm{P}}] - \frac{3}{2} i [ 1 + f_{\mathrm{e}}^2(\omega)] \Delta_{\mathrm{V}1}\biggr\}
\nonumber\\
&+& 
{\mathcal O}\biggl(\frac{m_{\mathrm{e}}}{m_{\mathrm{p}}}\biggr) + 
{\mathcal O}(f_{\mathrm{e}}^4).
\label{CexV1}
\end{eqnarray}
As anticipated, the source terms for the intensity and for the linear polarization 
have the first correction going as $f_{\mathrm{e}}^2(\omega)$ while
the source term for the circular polarization starts with $f_{\mathrm{e}}(\omega)$.
Higher order corrections to  Eqs. (\ref{CexI1}), (\ref{CexQ1}) 
and (\ref{CexV1})  can be computed, if needed recalling the results 
of Eqs. (\ref{LAM1})--(\ref{LAM3}) and of Eq. (\ref{LAMexp}).
Bearing in mind the results of Eqs. (\ref{CexI1}), (\ref{CexQ1}) 
and (\ref{CexV1}), to lowest order both in $(m_{\mathrm{e}}/m_{\mathrm{p}})$ and in $f_{\mathrm{e}}(\omega)$ 
 the following system of brightness perturbations can be obtained:
\begin{eqnarray}
&& \Delta_{\mathrm{I}}' + n^{i} \partial_{i}(\Delta_{\mathrm{I}} + \phi) + \epsilon' \Delta_{\mathrm{I}} = \psi' + 
\epsilon'\biggl[ \mu v_{\mathrm{b}} + \Delta_{\mathrm{I}0} - \frac{P_{2}(\mu) }{2} S_{\mathrm{P}}\biggr],
\label{LOI}\\
&&  \Delta_{\mathrm{P}}' + n^{i} \partial_{i}\Delta_{\mathrm{P}}  + \epsilon' \Delta_{\mathrm{P}} = 
\frac{3(1- \mu^2)\epsilon'}{4} S_{\mathrm{P}},
\label{LOP}\\
&& \Delta_{\mathrm{V}}' + n^{i} \partial_{i} \Delta_{\mathrm{V}} + \epsilon' \Delta_{\mathrm{V}} =\epsilon' P_{1}(\mu)
\biggl\{ \, f_{\mathrm{e}}(\omega) \biggl[ 2\Delta_{\mathrm{I}0} - S_{\mathrm{P}} \biggr]  - \frac{3}{4} i  \Delta_{\mathrm{V}1}\biggr\},
\label{LOV}
\end{eqnarray}
where all the corrections ${\mathcal O}(f_{\mathrm{e}}^2)$ have been neglected. Note that in Eq. (\ref{LOP}) 
$\Delta_{\mathrm{Q}}$ has been replaced by $\Delta_{\mathrm{P}}$, i.e. the brightness 
perturbation for the polarization degree $P = \sqrt{Q^2 + U^2}$. In equivalent terms, as customarily 
done, we could have chosen the frame where $\Delta_{\mathrm{U}} = 0$. Following the same notation, the source 
term $S_{\mathrm{P}}$ of Eq. (\ref{SP}) will become $S_{\mathrm{P}} =( \Delta_{\mathrm{I}2} + \Delta_{\mathrm{P}0} 
+ \Delta_{\mathrm{P}2})$.
If $f_{\mathrm{e}}(\omega) =0$ in Eq. (\ref{LOV}) the standard set of brightness perturbations is quickly recovered.
In this case the procedure will be to integrate the equations by assuming, for sufficiently 
early times, that the baryons are tightly coupled with the electrons implying that the baryon velocity is 
effectively equal to the dipole of the intensity, i.e. $v_{\mathrm{b}}  \simeq - 3 i \Delta_{\mathrm{I}1}$. This 
is, in a nutshell, the lowest order in the tight coupling expansion. To lowest order in the tight-coupling expansion 
the CMB is not polarized in the baryon rest frame, i.e. $\Delta_{\mathrm{I}0} \neq 0$ 
$\Delta_{\mathrm{I}1} \neq 0$ but $\Delta_{\mathrm{I}2} = \Delta_{\mathrm{P}2} = \Delta_{\mathrm{P}0} =0$. 
To first order in the tight coupling expansion (linear) polarization is generated and it is proportional, as expected, to the 
photon quadrupole which can be computed from the lowest order dipole. 
To summarize the approximations exploited so far we have that:
\begin{itemize}
\item{} the scattering matrix has been derived in the guiding centre 
approximation;
\item{} the brightness perturbations have been then expanded for $f_{\mathrm{e}}(\omega) < 1$;
\item{} the tight-coupling approximation is not invalidated by the new form of the evolution 
of the brightness perturbations.
\end{itemize}
Before giving the details on the line of sight solution of Eqs. (\ref{LOI}), (\ref{LOP}) and (\ref{LOV}) it is appropriate 
to pause a moment on the explicit numerical value of $f_{\mathrm{e}}(\omega)$ 
\begin{equation}
f_{\mathrm{e}}(\omega) = \frac{\omega_{\mathrm{Be}}}{\omega}= 2.8 \times 10^{-12} \biggl(\frac{B_{\mathrm{u}}}{\mathrm{nG}}\biggr)
 \biggl(\frac{\mathrm{GHz}}{\nu}\biggr) (z +1).
 \label{FE}
 \end{equation}
 For $z\simeq z_{\mathrm{rec}} \simeq 1091$ 
 (see e.g. \cite{WMAP5a,WMAP5b,WMAP5c}),  
 $f_{\mathrm{e}}(\omega)$ is of the order of $10^{-9}$ for 
 nG field strengths\footnote{The dependence upon the redshift comes about since 
the electron and proton masses break the Weyl invariance of the system (see appendix \ref{APPA} and, in particular, 
Eqs. (\ref{electronv})--(\ref{ionv}).}.  In  Eq. (\ref{FE}) $B_{\mathrm{u}}$ denotes the uniform component of the magnetic field, i.e. 
 we are assuming that the magnetic field is uniform since this is the simplest approximation in which 
 that heat transfer equations can be analyzed. The considerations reported here in the uniform field approximation 
 can be generalized to the case when the magnetic field is characterized by a given power spectrum. As already mentioned 
 at the end of section \ref{sec2} the uniform field approximation is more accurate in the present case than in the case of Faraday 
 rotation which will be left for future discussions.

The estimate of Eq. (\ref{FE}) can be further reduced by going to higher angular frequencies  where, typically, 
nearly all CMB experiments are operating\footnote{Just to have an idea the Planck explorer satellite is observing
 the microwave sky in nine frequency channels:
three frequency channels (i.e. $\nu= 30,\,44,\,70$ GHz) belong to the low frequency instrument (LFI);  six 
channels (i.e. $\nu= 100,\,143,\,217,\,353,\,545,\,857$ GHz) belong to the high 
frequency instrument (HFI). The five frequency channels 
of the WMAP experiment are centered at $23$, $33$, $41$, $61$ and $94$ in units of GHz. Neither 
WMAP nor Planck are sensitive to the circular polarizations.}. 
Even if $f_{\mathrm{e}}(\omega)$ can be rather small the question remains on the 
relative magnitude of the VV, VT and BB correlations and this will be one of the points discussed 
hereunder first for large angular scales (i.e. $\ell < 100$) and then for smaller angular scales when 
dissipative effects are important. 

\subsection{Line of sight solutions}
Equations (\ref{LOI}), (\ref{LOP}) and (\ref{LOV}) can be solved, formally, by integration along the line of sight and the result of this step can be written, in Fourier space, as 
\begin{eqnarray}
\Delta_{\mathrm{I}}(k,\mu,\tau_{0}) &=& \int_{0}^{\tau_{0}} e^{ - \epsilon(\tau,\tau_{0})}\,\,(\phi' + \psi') \,e^{- i \mu x} d\tau + 
\nonumber\\
&+& \int_{0}^{\tau_{0}} {\mathcal K}(\tau) \biggl[ \Delta_{\mathrm{I}0} + \mu v_{\mathrm{b}} - \frac{P_{2}(\mu)}{2} S_{\mathrm{P}}\biggr] e^{-i \mu x} \,d\tau,
\label{lineI}\\
\Delta_{\mathrm{P}}(k,\mu,\tau_{0}) &=& \frac{3(1 - \mu^2)}{4} \int_{0}^{\tau_{0}} {\mathcal K}(\tau) S_{\mathrm{P}} 
e^{- i \mu x} \, d\tau,
\label{lineP}\\
\Delta_{\mathrm{V}}(k,\mu, \omega,\tau_{0}) &=& \int_{0}^{\tau_{0}} {\mathcal K}(\tau) \, \mu\, \biggl\{ \, f_{\mathrm{e}}(\omega) \biggl[ 2\Delta_{\mathrm{I}0} - S_{\mathrm{P}} \biggr]  - \frac{3\, i}{4}   \Delta_{\mathrm{V}1}\biggr\} \,e^{-i \mu x} d\tau,
\label{lineV}
\end{eqnarray}
where, as usual, $x = k (\tau_{0} - \tau)$ and 
\begin{equation}
\epsilon(\tau, \tau_{0}) = \int_{\tau}^{\tau_{0}} x_{\mathrm{e}} \tilde{n}_{\mathrm{e}} a d\tau', \qquad 
{\mathcal K}(\tau) = \epsilon' e^{- \epsilon(\tau,\tau_{0})}.
\label{visibility}
\end{equation}
For large scales the visibility function ${\mathcal K}(\tau)$ can be taken as sharply peaked at the recombination time. For smaller 
angular scales the (approximately Gaussian) width is essential to obtain sound semi-analytical estimates.
Recalling the specific form of the lowest order Legendre polynomials \cite{abr1,abr2}
\begin{equation}
P_{0}(\mu) =1,\qquad P_{1}(\mu) = \mu,\qquad  P_{2}(\mu)= \frac{1}{2}(3 \mu^2 -1),
\end{equation}
 Eqs. (\ref{LOI})--(\ref{LOP}) and (\ref{LOV}) can be reduced to 
a hierarchy of coupled evolution equations for the various multipoles. 
Multiplying Eqs.  (\ref{LOI}), (\ref{LOP}) and (\ref{LOV}) by $P_{0}(\mu) = 1$ and integrating over $\mu$ between $-1$ and $1$, the following relations can be obtained
\begin{eqnarray}
&& \Delta_{{\rm I}0}' + k \Delta_{{\rm I}1} =  \psi',
\label{L01}\\
&& \Delta_{{\rm P}0} '+k \Delta_{{\rm P}1} = \frac{\epsilon'}{2} [ \Delta_{{\rm P}2} + \Delta_{{\rm I}2} - \Delta_{{\rm P}0} ],
\label{L02}\\
&& \Delta_{\mathrm{V}0}' + k \Delta_{\mathrm{V}1} = -
 \epsilon' \Delta_{\mathrm{V}0}. 
\label{L03}
\end{eqnarray}
If Eqs. (\ref{LOI})--(\ref{LOV}) are multiplied by  $P_{1}(\mu)$,
both at right and left-hand sides, 
the integration  over $\mu$ of the various terms implies:
\begin{eqnarray}
&& - \Delta_{{\rm I} 1}' - \frac{2}{3}k \Delta_{{\rm I}2} + \frac{k}{3} \Delta_{{\rm I}0} = - \frac{k}{3}  \phi + \epsilon' \biggl[ \Delta_{{\rm I} 1} + 
\frac{1}{3 i} v_{\rm b}\biggr],
\label{L11}\\
&& - \Delta_{{\rm P}1}' - \frac{2}{3} k \Delta_{{\rm P}2} + \frac{k}{3} \Delta_{{\rm P}0} = \epsilon' \Delta_{{\rm P} 1},
\label{L12}\\
&& \Delta_{\mathrm{V}1}' - \frac{2}{3} k \Delta_{\mathrm{V}2} + \frac{k}{3} 
\Delta_{\mathrm{V}0} = \epsilon' \biggl[ - \frac{3}{4} \Delta_{\mathrm{V}1} + 
\frac{i}{3} f_{\mathrm{e}} \biggl(2 \Delta_{\mathrm{I}0} - \Delta_{\mathrm{I}2} - 
\Delta_{\mathrm{P}2} - \Delta_{\mathrm{P}0}\biggr)\biggr].
\label{L13}
\end{eqnarray}
The same  procedure, using $P_{2}(\mu)$, leads to:
\begin{eqnarray}
&& - \Delta_{{\rm I} 2}' - \frac{3}{5} k \Delta_{{\rm I}3} + \frac{2}{5} k \Delta_{{\rm I} 1} = \epsilon'\biggl[ \frac{9}{10} \Delta_{{\rm I}2} - \frac{1}{10} (\Delta_{{\rm P}0} + 
\Delta_{{\rm P} 2} )\biggr],
\label{L21}\\
&&  - \Delta_{{\rm P} 2}' - \frac{3}{5} k \Delta_{{\rm P}3} + \frac{2}{5} k \Delta_{{\rm P} 1} = \epsilon'\biggl[ \frac{9}{10} \Delta_{{\rm P}2} - \frac{1}{10} (\Delta_{{\rm P}0} + 
\Delta_{{\rm I} 2} )\biggr],
\label{L22}\\
&& \Delta_{\mathrm{V}2}'   + \frac{k}{5} \biggl[ 3 
\Delta_{\mathrm{V}2} - 2 \Delta_{\mathrm{V}1}\biggr] = - \epsilon' 
\Delta_{\mathrm{V}2}.
\label{L23}
\end{eqnarray}
For $\ell\geq 3$ the hierarchy of the brightness can be determined in general terms by using 
the recurrence relation for the 
Legendre polynomials (see, e.g. \cite{abr1,abr2}); the result for $\ell\geq 3$ is:
\begin{eqnarray}
&&\Delta_{{\rm I}\ell}' + \epsilon' \Delta_{{\rm I}\ell} 
= \frac{k}{2 \ell + 1} [ \ell \Delta_{{\rm I}(\ell-1)} - (\ell + 1) \Delta_{{\rm I}(\ell + 1)}],
\nonumber\\
&& \Delta_{{\rm P}\ell}' + \epsilon' \Delta_{{\rm P}\ell} 
= \frac{k}{2 \ell + 1} [ \ell \Delta_{{\rm P}(\ell-1)} - (\ell + 1) \Delta_{{\rm P}(\ell + 1)}],
\nonumber\\
&&\Delta_{{\rm V}\ell}' + \epsilon' \Delta_{{\rm V}\ell} 
= \frac{k}{2 \ell + 1} [ \ell \Delta_{{\rm V}(\ell-1)} - (\ell + 1) \Delta_{{\rm V}(\ell + 1)}].
\end{eqnarray}
We are now ready to compute the evolution of the various terms to a given order in the tight-coupling expansion parameter $\tau_{\rm c} = |1/\epsilon'|$. 
After expanding the various moments of the brightness and of the baryon velocity in powers of $\tau_{\rm c}$, 
\begin{eqnarray}
&&\Delta_{{\rm I}\ell} = \overline{\Delta}_{{\rm I}\ell} + \tau_{\rm c} \delta_{{\rm I}\ell},
\qquad \Delta_{\mathrm{V}\ell} = \overline{\Delta}_{\mathrm{V}\ell} + \tau_{\rm c} \delta_{\mathrm{V}\ell},
\label{exp1}\\
&& \Delta_{{\rm P}\ell} = \overline{\Delta}_{{\rm P}\ell} + \tau_{\rm c} \delta_{{\rm P}\ell},
\qquad v_{\rm b} = \overline{v}_{\rm b} + \tau_{\rm c} \delta_{v_{\rm b}},
\label{exp2}
\end{eqnarray}
the obtained expressions can be inserted  into Eqs. (\ref{L01})--(\ref{L13}) and the evolution of the various moments of the brightness 
function can be found order by order in $\tau_{\mathrm{c}}$.
To zeroth order in the tight-coupling approximation, $\overline{v}_{b} = - 3 i  \overline{\Delta}_{{\rm I}1}$,
while Eqs. (\ref{L02}) and (\ref{L12}) lead, respectively, to
\begin{equation}
\overline{\Delta}_{{\rm P}0} = \overline{\Delta}_{{\rm I}2} + \overline{\Delta}_{{\rm P}2},
\qquad \overline{\Delta}_{{\rm P}1} =0,\qquad \overline{\Delta}_{\mathrm{V}0}=0.
\label{int1Q}
\end{equation}
Finally Eqs. (\ref{L21}) and (\ref{L22}) imply
\begin{equation}
9\overline{\Delta}_{{\rm I}2}  = \overline{\Delta}_{{\rm Q}0} + \overline{\Delta}_{{\rm Q}2},
\qquad 9\overline{\Delta}_{{\rm Q}2}  = \overline{\Delta}_{{\rm Q}0} + \overline{\Delta}_{{\rm I}2}, \qquad \overline{\Delta}_{\mathrm{V}1} = \frac{8}{9} \, i \, 
f_{\mathrm{e}}(\omega)\, \overline{\Delta}_{\mathrm{I}0}.
\label{int2Q}
\end{equation}
Taking together the four conditions expressed by Eqs. (\ref{int1Q}) and (\ref{int2Q}) we have, to zeroth order in the 
tight-coupling approximation:
\begin{eqnarray}
&& \overline{\Delta}_{{\rm Q}\ell} =0,\qquad \ell\geq 0,\qquad \overline{\Delta}_{{\rm I}\ell} =0,\qquad \ell \geq 2, 
\label{int3}\\
&& \overline{\Delta}_{\mathrm{V}\ell}=0, \qquad \ell \neq 1
\label{int3a}
\end{eqnarray}
Hence, to zeroth order in the tight-coupling, the relevant equations are 
\begin{eqnarray}
&& \overline{v}_{b} = - 3i \overline{\Delta}_{{\rm I}1},
\label{zerothorder1}\\
&& \overline{\Delta}_{{\rm I}0}' + k \overline{\Delta}_{{\rm I}1} =  \psi',
\label{zerothorder2}\\
&& \overline{\Delta}_{\mathrm{V}1} = \frac{8}{9} \, i \, 
f_{\mathrm{e}}(\omega)\, \overline{\Delta}_{\mathrm{I}0}.
\label{zerothorder3}
\end{eqnarray}

Even if to zeroth order in the tight coupling expansion we have that 
the linear polarization is absent. To get a non-vanishing linear polarization we have to go to first-order where the monopole 
and the dipole of the linear polarization are proportional 
to the quadrupole of the intensity; at the same order in the 
perturbative expansion the a non-vanishing quadrupole 
of the circular polarization is also generated. Indeed, recalling 
Eqs. (\ref{exp1}) and (\ref{exp2}), the first-order results 
can be written as 
\begin{eqnarray}
&& \delta_{{\rm Q}0} = \frac{5}{4} \delta_{{\rm I}2},\qquad 
 \delta_{{\rm Q}2} = \frac{1}{4} \delta_{{\rm I}2},
\label{Cond1}\\
&& \delta_{{\rm I}2} = \frac{8}{15} k \overline{\Delta}_{{\rm I}1},
\qquad \delta_{{\rm V}2} = \frac{2}{5} k \overline{\Delta}_{{\rm V}1}.
\label{Cond2}
\end{eqnarray}
From Eqs. (\ref{Cond1}) and (\ref{Cond2}) the line of sight solution 
of Eq. (\ref{lineV}) can be written as
\begin{equation}
\Delta(k,\mu, \omega, \tau_{0}) = \int_{0}^{\tau_{0}} {\mathcal K}(\tau) \mu \biggl\{ \frac{8}{3} f_{\mathrm{e}}(\omega) \overline{\Delta}_{\mathrm{I}0}(k, \tau) - \frac{f_{\mathrm{e}}(\omega)}{3} 
(k\tau_{\mathrm{c}}) \overline{\Delta}_{\mathrm{I}1} \biggr\} e^{- i \mu x} d\tau,
\label{LSV1}
\end{equation}
where Eqs. (\ref{zerothorder1}) and (\ref{zerothorder2})  have been also used. The result of Eq. (\ref{LSV1}) has been used in 
\cite{maxsh} and the present discussion corroborate and extends those results.

In the sudden decoupling approximation 
the visibility function becomes effectively a Dirac delta 
function and the second term (proportional to the 
dipole of the intensity) can be neglected in comparison 
with the monopole whose evolution equation 
can be written as 
\begin{eqnarray}
\overline{\Delta}_{\mathrm{I}0}'' + \frac{{\mathcal H} R_{\mathrm{b}}}{R_{\mathrm{b}} + 1} \overline{\Delta}_{\mathrm{I}0}' + 
\frac{k^2}{3 ( R_{\mathrm{b}} + 1)} \overline{\Delta}_{\mathrm{I}0} = \Sigma_{\psi} + \Sigma_{\mathrm{B}},
\label{eff2}
\end{eqnarray}
where $R_{\mathrm{b}} = (3/4) \rho_{\mathrm{b}}/\rho_{\gamma}$ is the baryon to photon ratio and where 
\begin{eqnarray}
\Sigma_{\psi} &=& \psi'' + \frac{{\mathcal H} R_{\mathrm{b}}}{ R_{\mathrm{b}} + 1} \psi'  - \frac{k^2}{3} \phi
\label{Seff1}\\
\Sigma_{\mathrm{B}} &=&  \frac{k^2}{12(R_{\mathrm{b}} +1)} ( 4 \sigma_{\mathrm{B}} - \Omega_{\mathrm{B}}).
\label{Seff2}
\end{eqnarray}
 Equation (\ref{Seff2}) accounts for the 
presence of an inhomogenous magnetic field (see \cite{max2,max6,max7} for a definition) but in the estimates we 
are going to present here, the inhomogeneities 
stemming from the magnetic field itself will be neglected for consistency with the uniform field approximation.  

It is finally appropriate to remark that it is possible to go improve on the accuracy of the 
line of sight solutions for the V-mode polarization. For this purpose the technique developed in Ref. \cite{sp2} will be extended to the line of sight solution 
of the V-mode polarization. To be specific consider, again, the system of Eqs. 
(\ref{lineI}), (\ref{lineP}) and (\ref{lineV}). In particular we shall be interested in improving on Eq. 
(\ref{LSV1}) which has been derived from Eq. (\ref{lineV}). Instead of using the lowest order 
tight-coupling solution for the linear polarization source it is possible to resum the perturbative 
expansion by solving an effective evolution equation for $S_{\mathrm{P}}$. Recall, for this purpose, 
 that the conformal time derivative of  $S_{\mathrm{P}}$, i.e. 
$S_{\mathrm{P}}' = ( \Delta_{\mathrm{I}2}' + \Delta_{\mathrm{P}0}' 
+ \Delta_{\mathrm{P}2}')$ can be expressed as the sum of the evolution equations 
of the separate multipoles. In particuylar $\Delta_{\mathrm{P}0}'$ can be expressed from Eq.  (\ref{L02}), 
$\Delta_{\mathrm{I}2}'$ can be expressed from Eq. (\ref{L21}) and $\Delta_{\mathrm{P}2}'$ from Eq. (\ref{L22}).
Summing up the various contributions 
and rearranging the relevant terms an effective evolution equation for $S_{\mathrm{P}}$ can be obtained and it is 
 \begin{equation}
 S_{\rm P}' + \frac{3}{10} \epsilon' S_{\rm P} = k \biggl[ \frac{2}{5} \Delta_{{\rm I}1} - 
 \frac{3}{5}\biggl( \Delta_{{\rm P}1}+ \Delta_{{\rm P}3} + \Delta_{{\rm I}3}\biggr)\biggr],
\label{SP4}
 \end{equation}
which can be solved in different ways. For instance, if only the intensity dipole is kept, at the right hand side of Eq. 
(\ref{SP4}) the result for $S_{\mathrm{P}}$ is 
\begin{equation}
S_{\rm P}(k,\tau) =\frac{2}{5} k e^{ 3 \epsilon(\tau,\tau_0)/10} \int_{0}^{\tau}  d \tau' \overline{\Delta}_{{\rm I}1}(k,\tau') e^{-3 \epsilon(\tau',\tau_{0})/10}.
\label{SP6}
\end{equation}
If we now insert Eq. (\ref{SP6}) into Eq. (\ref{lineV}) the integrals can be performed, up to some point, 
with analytic techniques. This result improves then Eq. (\ref{LSV1}).  

The considerations developed so far suggest the following physical picture. To lowest order in the 
tight-coupling expansion the presence of a magnetic field produces a dipole 
of the circular polarization. The dipole of the V-mode can be fed back into the line of sight solution to 
obtain the higher multipoles in full analogy with what is customarily done in Boltzmann solvers. While the circular polarization is building up from the monopole of the 
intensity, the dipole of the intensity sources, to first-order in the tight coupling expansion,
the linear polarization and, in particular, $\Delta_{\mathrm{P}0}$ and $\Delta_{\mathrm{P}1}$.
The interesting aspect of this analysis is that, indeed, to lowest order in the 
tight coupling approximation the CMB is circularly polarized if a pre-decoupling magnetic field is 
present. The linear polarization is generated to first-order in the tight-coupling expansion.
At the level of the amplitude, as it will be shown, the angular power spectrum 
of the V-mode is always smaller than the E-mode spectrum which arises directly from 
$\Delta_{\mathrm{P}}$. The reason for this occurrence stems from the specific value of 
$f_{\mathrm{e}}(\omega)$.
 
\subsection{Large-scale limit}
\begin{figure}[!ht]
\centering
\includegraphics[height=6cm]{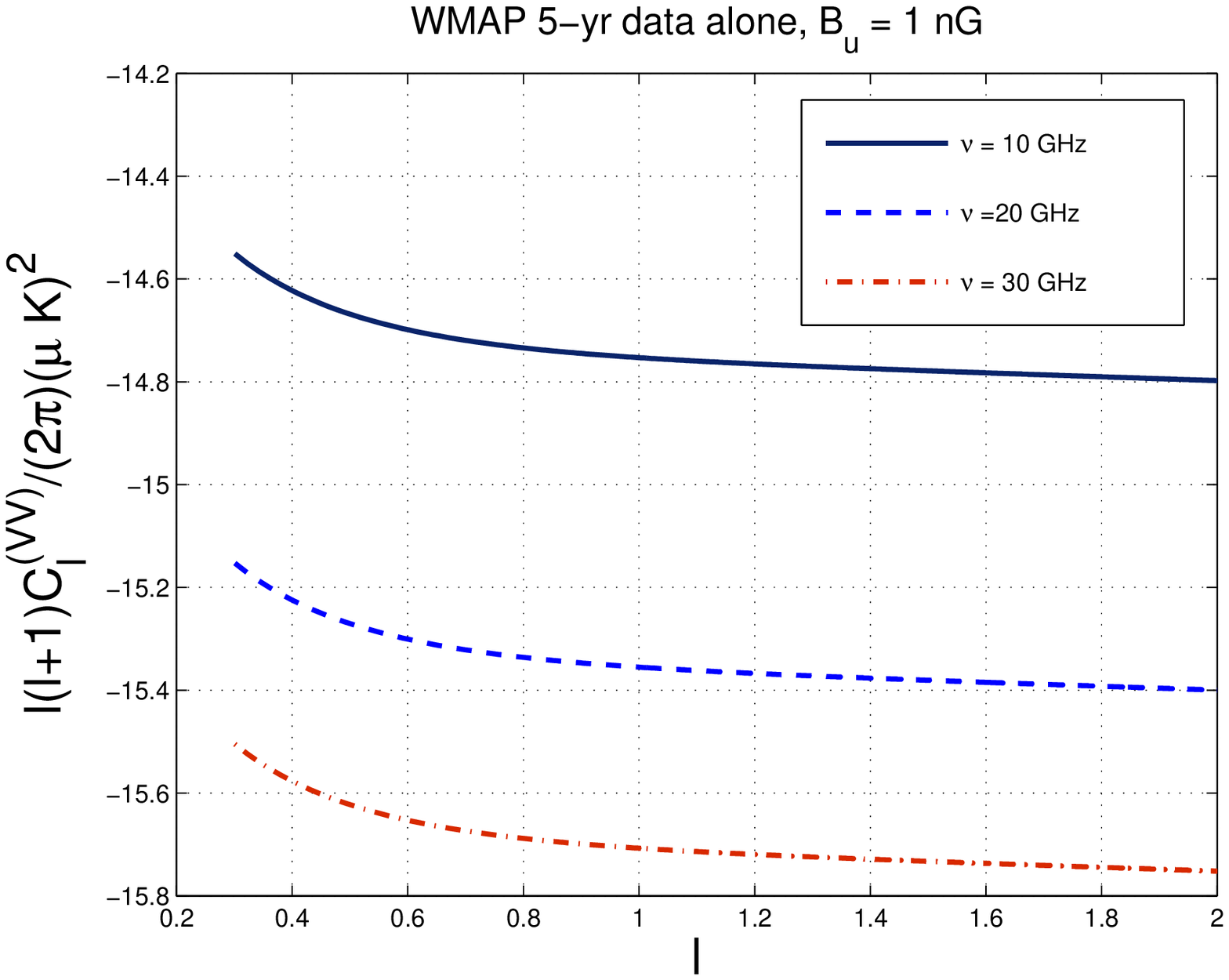}
\includegraphics[height=6cm]{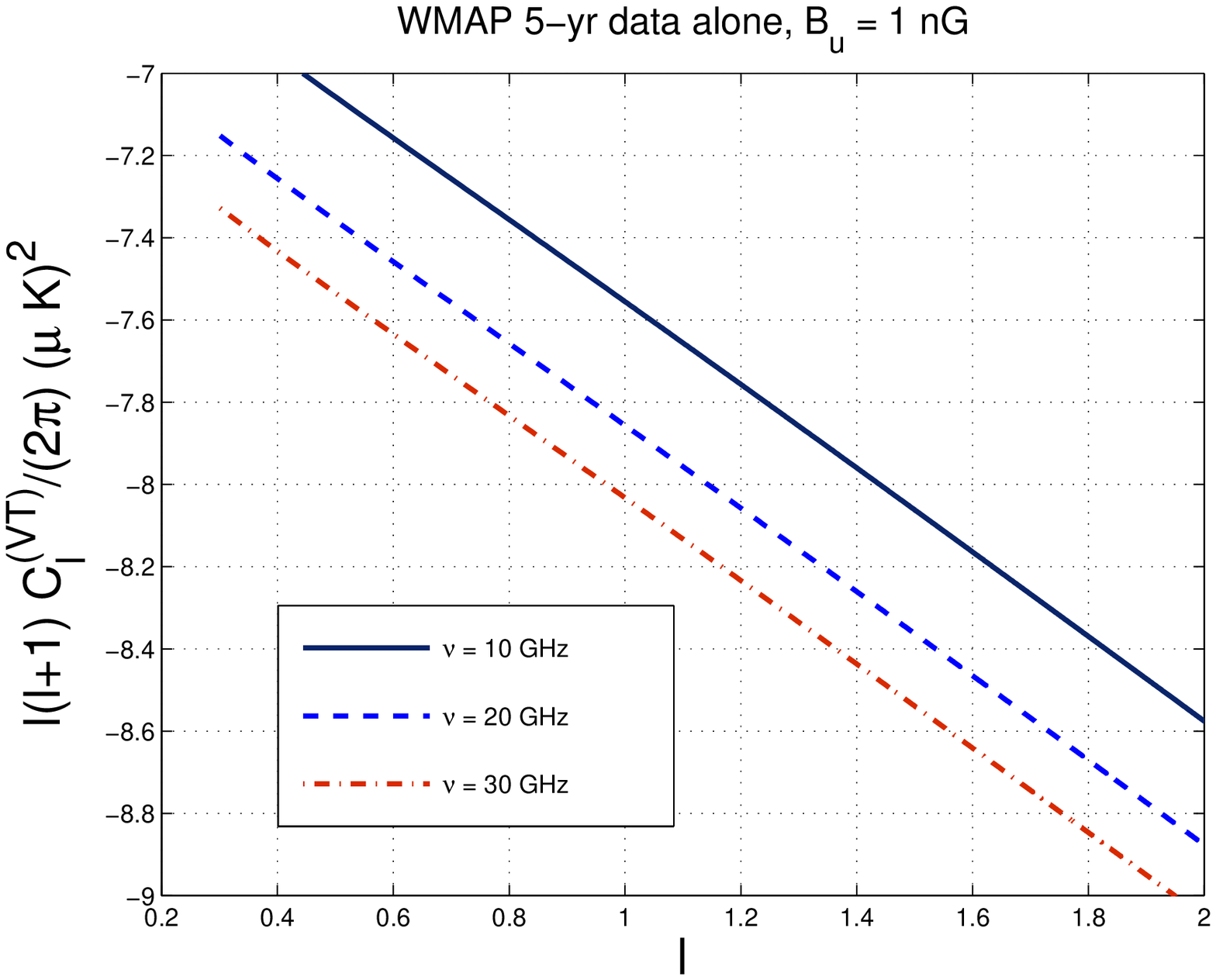}
\caption[a]{The VV and VT angular power spectra are illustrated for large angular scales and for different frequencies at a fixed value 
of the magnetic field intensity. In both plots a double logarithmic scale has been 
employed, i.e. on both axes we plot the common logarithms of the corresponding 
quantity.}
\label{figure1}      
\end{figure}

\begin{figure}[!ht]
\centering
\includegraphics[height=6cm]{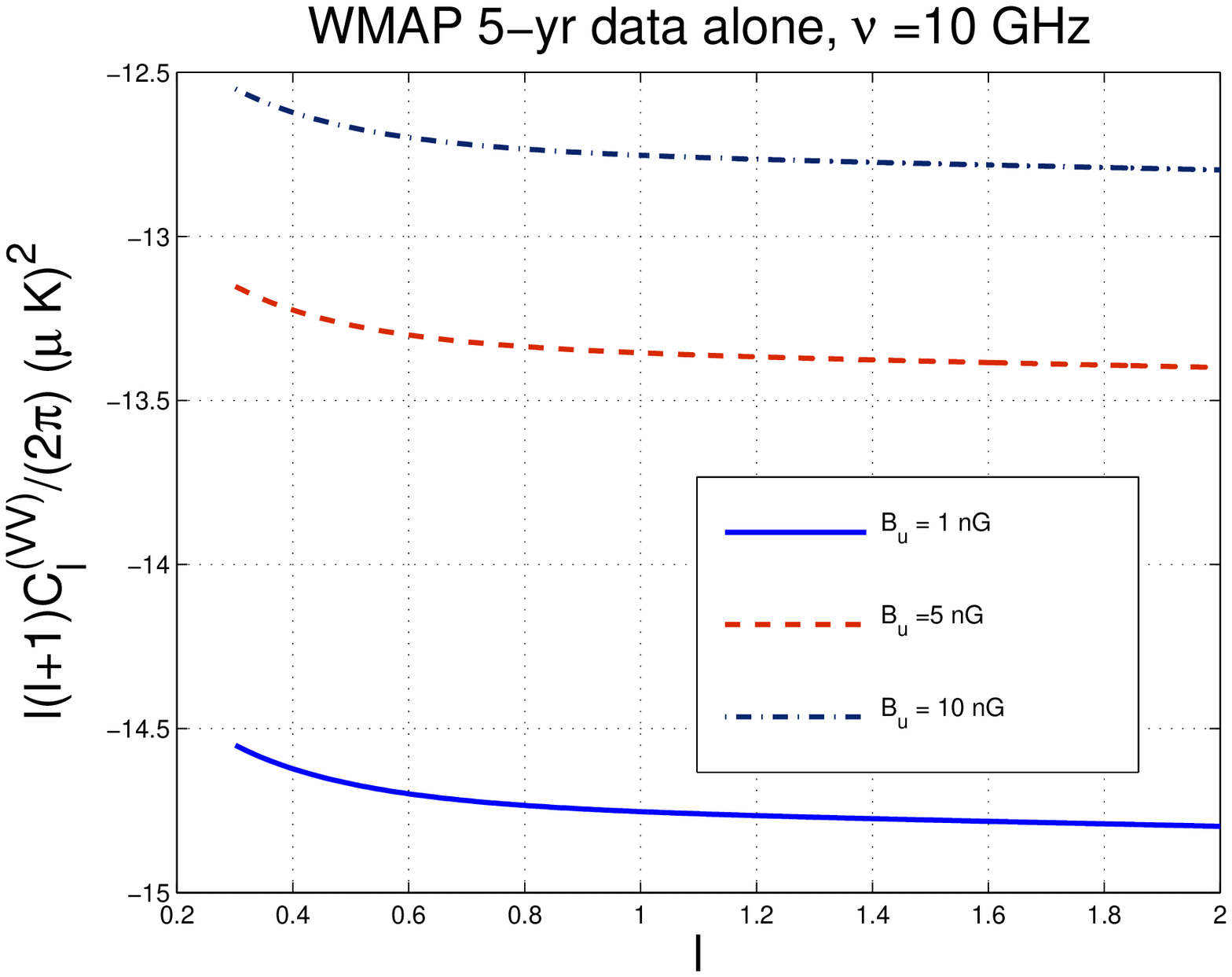}
\includegraphics[height=6cm]{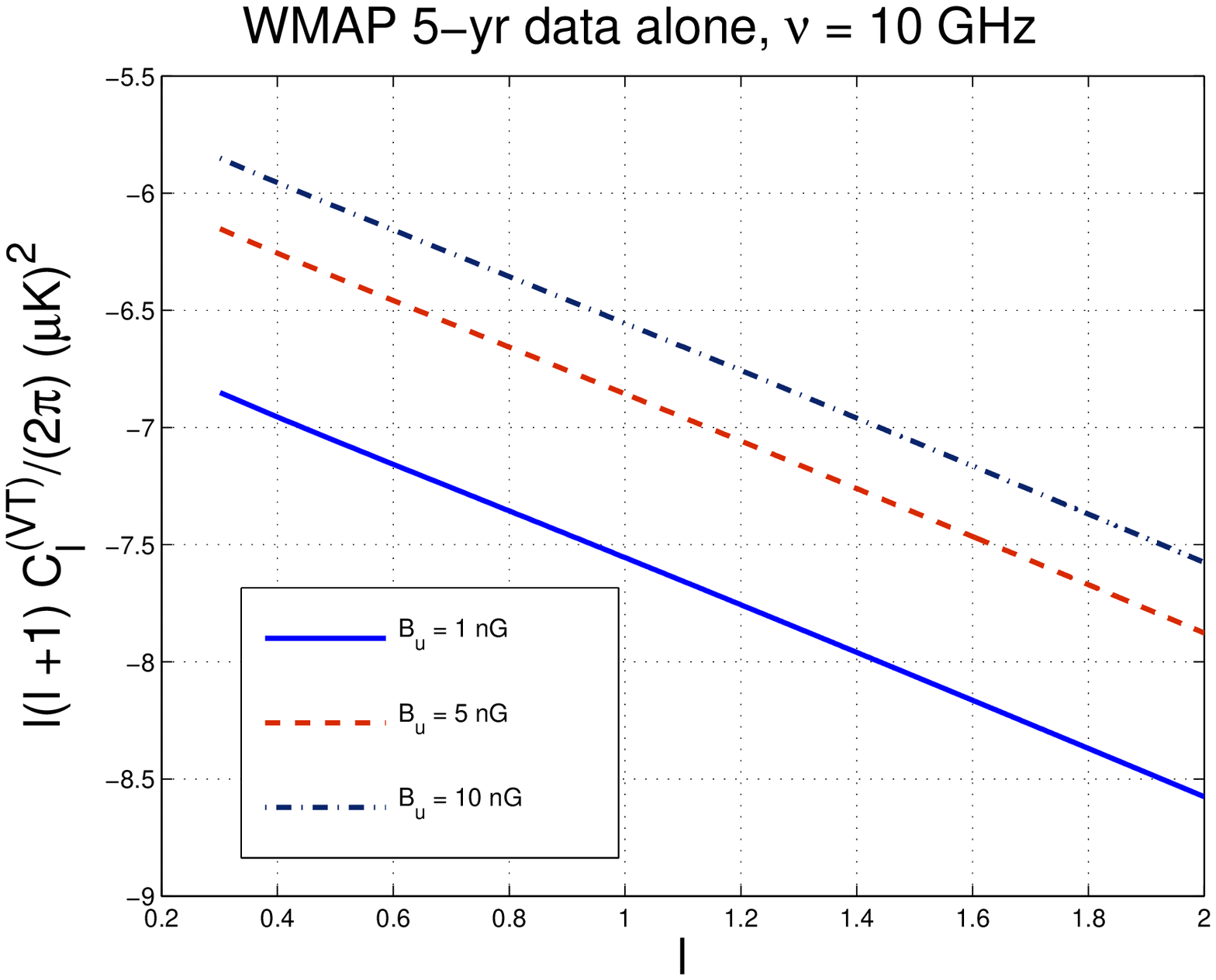}
\caption[a]{The VV and VT angular power spectra are illustrated for large angular scales and for different values of the magnetic field intensity at a fixed value 
of the frequency. As in Fig. \ref{figure1} a double logarithmic scale 
has been employed in both plots of the present figure.}
\label{figure2}      
\end{figure}

For large angular scales the  circular polarization will then be given by
\begin{equation}
\Delta_{\mathrm{V}}(k,\mu,\tau_{0}) = \frac{8}{3}  \int_{0}^{\tau_{0}} {\mathcal K}(\tau) f_{\mathrm{e}}(\omega)
e^{- i \mu k (\tau_{0} - \tau)} \, \mu\, \overline{\Delta}_{\mathrm{I}0}(k,\tau) \, d\tau,
\label{LS1}
\end{equation}
where $\overline{\Delta}_{\mathrm{I}0}$ can be determined from Eqs. (\ref{zerothorder1})--(\ref{zerothorder3}) and (\ref{eff2}). The terms arising in $S_{\mathrm{P}}$ have been neglected since they vanish to lowest order in the tight coupling expansion.  By assuming that ${\mathcal K}(\tau)$ is a Dirac delta function centered at recombination
(sudden decoupling approximation) we shall have that 
\begin{equation}
\Delta_{\mathrm{V}}(k,\mu,\tau_{0},\omega) = \frac{8}{3} f_{\mathrm{e}}(\omega) e^{- i \mu x}\,
 \mu \,\overline{\Delta}_{\mathrm{I}0}(k,\tau_{\mathrm{rec}}).
\end{equation}
Let us now compute the angular power spectrum of the circular polarization. 
\begin{equation}
\Delta_{\mathrm{V}}(\hat{n},\tau_{0}) = \sum_{\ell,\, m} a_{\ell\,m}^{(\mathrm{V})} Y_{\ell\,m}(\hat{n}),
\end{equation}
thus we will also have that 
\begin{equation}
a_{\ell m}^{(\mathrm{V})} = \int d\hat{n} \Delta_{\mathrm{V}}(\hat{n},\tau_{0})Y^{*}_{\ell\,m}(\hat{n})
= \frac{1}{(2\pi)^{3/2}} \int d\hat{n} \int d^{3} k \Delta_{\mathrm{V}}(k,\mu,\tau_{0}) \, 
Y_{\ell\, m}^{*}(\hat{n})
\end{equation}
Recalling the explicit expression of 
$\Delta_{\mathrm{V}}(k,\mu,\tau_{0})$
\begin{eqnarray}
a_{\ell\, m}^{(\mathrm{V})} &=& \frac{8 f_{\mathrm{e}}(\omega)}{3 (2\pi)^{3/2}} \int d\hat{n} \int d^{3} k
Y_{\ell\, m}^{*}(\hat{n}) \mu e^{- i \mu x} \Delta_{\mathrm{I}0}(k,\tau_{\mathrm{rec}})
 \nonumber\\
&=& \frac{8 i\,f_{\mathrm{e}}(\omega)}{3(2\pi)^{3/2}} \int d\hat{n} \int d^{3} k
Y_{\ell\, m}^{*}(\hat{n}) \frac{d}{dx}\biggl(e^{- i \mu x}\biggr) \Delta_{\mathrm{I}0}(k,\tau_{\mathrm{rec}}).
\end{eqnarray}
The integration over $d\hat{n} = \sin{\vartheta} d\varphi d \vartheta = - d\mu d\varphi$ can be performed in explicit terms since 
$e^{- i \mu x}$ can be expanded in Rayleigh series and the 
final result is  
\begin{equation}
a_{\ell\, m}^{(\mathrm{V})} 
=- \frac{8}{3}\delta_{m 0} \frac{(-i)^{\ell + 1}}{(2\pi)^{3/2}} f_{\mathrm{e}} \sqrt{4\pi} \sqrt{2\ell +1} \int d^{3} k \biggl(\frac{d j_{\ell}}{d x}\biggr) 
\Delta_{\mathrm{I}0}(k, \tau_{\mathrm{rec}}).
\end{equation}
The angular power spectrum of the circular polarization can then be written as 
\begin{eqnarray}
C_{\ell}^{(\mathrm{VV})}(\omega) &=& \frac{1}{2\ell +1} \sum_{m = - \ell}^{\ell} \langle a_{\ell\, m}^{(V)*}
a_{\ell\, m}^{(V)}  \rangle 
\nonumber\\
&=& \frac{256\pi}{9} f_{\mathrm{e}}^2(\omega) \int_{0}^{\infty} \frac{d k}{k} \frac{k^3}{2\pi^2} \,
\biggl(\frac{d j_{\ell}}{d x}\biggr)^2\,
|\Delta_{\mathrm{I}0}(k,\tau_{\mathrm{rec}})|^2,
\end{eqnarray}
where the dependence on the angular frequency $\omega$ has been explicitly included in the expression 
of the angular power spectrum. 
The  large-scale estimate of $\overline{\Delta}_{\mathrm{I}0}(k,\tau_{\mathrm{rec}})$ follows 
from Eq. (\ref{L01}) by neglecting the dipole which is negligible 
for those wavelengths which are still larger than the Hubble radius 
around the redshift of recombination:
\begin{equation}
\overline{\Delta}_{\mathrm{I}0}(k,\tau_{\mathrm{rec}})=\overline{\Delta}_{\mathrm{I}0}(k,\tau_{*}) 
 + \psi(k,\tau_{\mathrm{rec}}) - \psi_{*}(k),
\end{equation}
where, by definition, $\psi(k,\tau_{*}) = \psi_{*}(k)$ and $\phi(k,\tau_{*}) = \phi_{*}(k)$ are the values of the metric fluctuations 
at $\tau_{*} \ll \tau_{\mathrm{eq}}$. For $\tau \simeq \tau_{*}$ 
$k \tau_{*} \ll1$ and, 
in the minimal $\Lambda$CDM scenario, the initial conditions 
are (predominantly) adiabatic, i.e.  
\begin{equation}
\delta_{\gamma}(k,\tau_{*}) = - 2 \phi_{*}(k), \qquad \psi(k,\tau_{*}) = \psi_{*}(k)
\qquad {\mathcal R}_{*}(k) = - \psi_{*}(k) - \frac{\phi_{*}(k)}{2},
\end{equation}
implying
\begin{equation}
\Delta_{\mathrm{I}0}(k,\tau_{\mathrm{rec}})= \frac{ 2 ( R_{\nu} + 15)}{5 ( 4 R_{\nu} + 15)} {\mathcal R}_{*}(k),\qquad  \psi_{*}(k) = \biggl( 1 + \frac{2}{5} R_{\nu} \biggr) \phi_{*}(k),
\end{equation}
where $R_{\nu}$ is the fractional contribution of the massless 
neutrinos to the radiation background\footnote{Neutrinos 
are taken to be massless for consistency with 
$\Lambda$CDM paradigm even if the effect of the masses 
could be included without appreciable changes in the 
forthcoming numerical estimates.}. 
The angular power spectrum is 
\begin{eqnarray}
C_{\ell}^{(\mathrm{VV})}(\omega) &=& \frac{512 \pi^2}{225} \biggl(\frac{R_{\nu} + 15}{4 R_{\nu} + 15}\biggr)^2 f_{\mathrm{e}}^2(\omega)  {\mathcal A}_{{\mathcal R}} \biggl(\frac{k_{0}}{k_{\mathrm{p}}}\biggr)^{n_{\mathrm{s}}-1} {\mathcal I}^{(\mathrm{VV})}(\ell, n_{\mathrm{s}}),
\nonumber\\
{\mathcal I}^{(\mathrm{VV})}(\ell, n_{\mathrm{s}}) &=& \frac{2\ell (\ell + 1)}{\pi} \int_{0}^{\infty} d x \,\,x^{n_{\mathrm{s}} -2} \biggl(\frac{d j_{\ell}}{dx}\biggr)^2.
\nonumber
\end{eqnarray}
The derivative of the spherical Bessel function can be expressed in terms of the 
appropriate recurrence relations, namely \cite{abr1,abr2}:
\begin{equation}
\frac{d j_{\ell}}{d x} = \frac{\ell}{x} j_{\ell}(x) - j_{\ell +1}(x).
\nonumber
\end{equation}
The integral becomes then:
\begin{equation}
{\mathcal I}^{(\mathrm{VV})}(\ell, n_{\mathrm{s}}) = \ell (\ell +1) \int_{0}^{\infty}\, dx\, x^{n_{\mathrm{s}} -5} 
\biggl[ \ell^2 J^2_{\ell +1/2}(x) + x^2 J_{\ell +3/2}^2(x) - 2 \ell x J_{\ell+1/2}(x) J_{\ell +3/2}(x) \biggr],
\label{INTVV}
\end{equation}
which can be explicitly computed \cite{abr1,abr2}.
The final result for the angular power spectrum can therefore be written  as
\begin{equation}
\frac{\ell (\ell +1)}{2\pi} C_{\ell}^{(\mathrm{VV})}(\omega) =  \frac{256 \pi}{225} \biggl(\frac{R_{\nu} + 15}{4 R_{\nu} + 15}\biggr)^2 f_{\mathrm{e}}^2(\omega) {\mathcal A}_{{\mathcal R}} \biggl(\frac{k_{0}}{k_{\mathrm{p}}}\biggr)^{n_{\mathrm{s}}-1}  {\mathcal I}^{(\mathrm{VV})}(\ell, n_{\mathrm{s}}),
\label{LSVV1}
\end{equation}
where 
\begin{equation}
{\mathcal I}_{\ell}^{(\mathrm{VV})}(n_{\mathrm{s}}) = 
\frac{\ell(\ell +1)[ 4 \ell (\ell +1) - (n_{\mathrm{s}} -1) (n_{\mathrm{s}}-2) (n_{\mathrm{s}}-4)] \Gamma(3 -n_{\mathrm{s}}) \Gamma\biggl(\ell - \frac{3}{2} + \frac{n_{\mathrm{s}}}{2}\biggr)}{2^{6 - n_{\mathrm{s}}} \Gamma\biggl(2 - \frac{n_{\mathrm{s}}}{2}\biggr)
\Gamma\biggl(3 - \frac{n_{\mathrm{s}}}{2}\biggr) \Gamma\biggl(\frac{7}{2} +\ell - 
\frac{n_{\mathrm{s}}}{2} \biggr)}.
\nonumber
\end{equation}
Following the same technique we can estimate the cross-correlation between temperature and polarization as:
\begin{equation}
\frac{\ell (\ell +1)}{2\pi} C_{\ell}^{(\mathrm{VT})}(\omega) = 
\frac{16\pi}{75} f_{\mathrm{e}}(\omega) \frac{(R_{\nu} + 15) ( 2 R_{\nu} - 15)}{(4 R_{\nu} + 15)^2} 
\biggl(\frac{k_{0}}{k_{\mathrm{p}}}\biggr)^{n_{\mathrm{s}} -1} \,
{\mathcal I}^{(\mathrm{VT})}(\ell, n_{\mathrm{s}}),
\end{equation}
where 
\begin{equation}
{\mathcal I}^{(\mathrm{VT})}(\ell, n_{\mathrm{s}}) = \ell (\ell +1) \int_{0}^{\infty}  dx \, x^{n_{\mathrm{s}} - 4} \biggl[ \ell 
\,J_{\ell +1/2}^2(x) - x J_{\ell +1/2}(x) J_{\ell +3/2}(x)\biggr].
\end{equation}
By integrating the above expression we have that:
\begin{equation}
{\mathcal I}^{(\mathrm{VT})}(\ell, n_{\mathrm{s}}) = \frac{\ell (\ell +1)(2-n_{\mathrm{s}}) \Gamma\biggl(2-\frac{n_{\mathrm{s}}}{2}\biggr) 
\Gamma\biggl(\ell+\frac{n_{\mathrm{s}}}{2}-1\biggr)}{4 \sqrt{\pi } \Gamma\biggl(\frac{5}{2}-\frac{n_{\mathrm{s}}}{2}\biggr)
 \Gamma\biggl(\ell-\frac{n_{\mathrm{s}}}{2}+3\biggr)}.
 \nonumber
\end{equation}
In Fig. \ref{figure1}  the VV and the VT angular power spectra are illustrated  for large angular scales (i.e. $\ell <100$).  Both in Fig. \ref{figure1}
and in Fig. \ref{figure2}  a double logarithmic scale has been used.  In Fig. \ref{figure1} the uniform magnetic field intensity is fixed while 
the frequency ranges between $10$ GHz and $30$ GHz. Note that $30$ GHz corresponds to the lower frequency band of the
 Planck explorer satellite which is unfortunately not sensitive to the circular polarization. 
In Fig. \ref{figure1} the cosmological parameters have been fixed as 
\begin{equation}
( \Omega_{\mathrm{b}}, \, \Omega_{\mathrm{c}}, \Omega_{\mathrm{de}},\, h_{0},\,n_{\mathrm{s}},\, \epsilon_{\mathrm{re}}) \equiv 
(0.0441,\, 0.214,\, 0.742,\,0.719,\, 0.963,\,0.087),
\label{Par1}
\end{equation}
corresponding to the best fit of the WMAP data alone in the light of the vanilla $\Lambda$CDM.
In Fig. \ref{figure2} the frequency is fixed to $10$ GHz while the magnetic field intensity changes from $1$ to $10$ nG.
The values of the magnetic field intensity are motivated by a recent analysis \cite{max2} where using the TE and TT correlations 
the parameters of a putative magnetized background are scrutinized. According to \cite{max1,max2} nG magnetic fields and 
blue magnetic spectral indices are fully compatible with the measured values of the TT and TE angular power spectra. 

The results of Fig. \ref{figure1} and of Fig. \ref{figure2} show already that the 
 values of the VT correlations 
are close to the magnitude of the BB angular power spectra from gravitational lensing as well as to the 
BB angular power spectra expected from the tensor modes of the geometry. It is important to remind here 
that the VV and  VT angular power spectra are a direct consequence of, both, the weakly magnetized plasma 
and the adiabatic mode of curvature perturbations. Absent one of these 
two essential ingredients the net result would vanish.  The considerations reported so far 
complement some of the results already reported in \cite{maxsh}.
The large-scale (analytical) estimates will now be corroborated by the numerical results at smaller angular scales. 
\subsection{Small-scale limit}
The visibility function vanishes  for  $\tau \gg \tau_{\mathrm{rec}}$
and has a maximum around recombination, i.e. when  
\begin{equation}
\epsilon'' + \epsilon'^2 =0, \qquad \frac{d \epsilon}{d \tau} = - x_{\mathrm{e}}(\tau) \tilde{n}_{\mathrm{e}}(\tau) \sigma_{\mathrm{Th}} a(\tau) \equiv - \epsilon'.
\label{rot2c}
\end{equation}
The second expression in Eq. (\ref{rot2c}) clarifies that a minus sign appears in the time derivative of $\epsilon(\tau,\tau_{0})$ since $\tau$ appears in the lower limit of integration.  When the finite thickness effects of the last scattering surface 
are taken into account the visibility function can be approximated by a Gaussian profile centered at $\tau_{\mathrm{rec}}$, i.e. 
\begin{eqnarray}
{\mathcal K}(\tau) &=& {\mathcal N}(\sigma_{\mathrm{rec}}) e^{- \frac{(\tau - \tau_{\mathrm{rec}})^2}{2 \sigma_{\mathrm{rec}}^2}},\qquad \int_{0}^{\tau_{0}} {\mathcal K}(\tau) d\tau = 1, 
\label{rot5}\\
{\mathcal N}(\sigma_{\mathrm{rec}})&=& \sqrt{\frac{2}{\pi}} \frac{1}{\sigma_{\mathrm{rec}}} 
\biggl[ \mathrm{erf}\biggl( \frac{\tau_{0} - \tau_{\mathrm{rec}}}{\sqrt{2} \sigma_{\mathrm{rec}}} \biggr)+   \mathrm{erf}\biggl( \frac{ \tau_{\mathrm{rec}}}{\sqrt{2} \sigma_{\mathrm{rec}}} \biggr)\biggr]^{-1}, 
\label{rot6}\\
\mathrm{erf}(z) &=& \frac{2}{\sqrt{\pi}} \int_{0}^{z} e^{- t^2} dt.
\end{eqnarray}
The overall normalization ${\mathcal N}(\sigma_{\mathrm{rec}})$ has been 
chosen in such a way that the integral of ${\mathcal K}(\tau)$ is normalized to 
$1$: the visibility function is nothing but the probability that a photon last scatters between $\tau$ and $\tau + d\tau$. Equation
 (\ref{rot6}) simplifies when $\tau_{0} \gg \tau_{\mathrm{rec}}$ and 
 $ \tau_{0} \gg \sigma_{\mathrm{rec}}$, since, in this limit, the error 
 functions go  to a constant and ${\mathcal N}(\sigma_{\mathrm{rec}}) \to\sigma_{\mathrm{rec}}^{-1}\, \sqrt{2/\pi}$. In the latter limit, the thickness of the last scattering surface, i.e. $\sigma_{\mathrm{rec}}$, is of the order of $\tau_{\mathrm{rec}}$. The Gaussian approximation 
 for the visibility function has a 
 long history (see, e.g. \cite{zeld1,wyse,pav1,pav2}).
 The WMAP data suggest a thickness 
(in redshift space) $\Delta z_{\mathrm{rec}} \simeq 195 \pm 2$  which would imply 
that $\sigma_{\mathrm{rec}}$, in units of the (comoving) angular diameter distance to recombination, 
can be estimated as  $\sigma_{\mathrm{rec}}/\tau_{0} \simeq 1.43 \times 10^{-3}$. 
By using the finite width of the visibility function we have that  
\begin{equation}
a_{\ell\, m}^{(\mathrm{V})} =- \delta_{m 0} \frac{8(-i)^{\ell + 1}}{3(2\pi)^{3/2}} f_{\mathrm{e}} \sqrt{4\pi} \sqrt{2\ell +1} \int d^{3} k \, \biggl(\frac{dj_{\ell}}{dx} \biggr) \, e^{- \frac{k^2}{k_{\mathrm{t}}^2}}\,\mu\,\overline{\Delta}_{\mathrm{I}0}(k, \tau_{\mathrm{rec}}) e^{- \epsilon_{\mathrm{re}}}.
\end{equation}
where $k_{\mathrm{t}} = (\sqrt{6}/\sigma_{\mathrm{rec}})$, $x = k(\tau_{0} - \tau_{\mathrm{rec}})$. The visibility function, more realistically, has also a second peak 
 at the reionization epoch (i.e. for $z_{\mathrm{re}} = 11\pm 1.4$). Also in this case 
 the visibility function can be approximated with a Gaussian profile centered, 
 this time, around $\tau_{\mathrm{re}}$ and this consideration introduces 
 a suppression going as $e^{- \epsilon_{\mathrm{re}}}$.
 In the limit $\tau_{0} \gg \tau_{\mathrm{rec}}$ and 
 $ \tau_{0} \gg \sigma_{\mathrm{rec}}$ the integral to be computed is, therefore, 
\begin{equation}
\frac{\ell (\ell +1)}{2\pi} C_{\ell}^{(\mathrm{VV})}(\omega) = 8 f_{\mathrm{e}}^2(\omega) \int_{0}^{\infty} \, e^{ - 2 \frac{k^2}{k_{\mathrm{t}}^2}} \frac{k^3}{2 \pi^2 } |\Delta_{\mathrm{I}0}(k,\tau_{\mathrm{rec}})|^2 \, (\ell + 1) \ell \biggl(\frac{d j_{\ell}}{dx}\biggr)^2.
\end{equation}  
Recalling that 
\begin{equation}
\biggl(\frac{d j_{\ell}}{d x}\biggr)^2 = \biggl[ 1 - \frac{\ell(\ell +1)}{x^2}\biggr] j_{\ell}^2(x) + \frac{1}{2 x} \frac{d^2}{dx^2} [ x j_{\ell}^2(x)]
\end{equation}
the Bessel functions can be estimated in the limit of very large 
multipoles. A standard calculation then leads to 
\begin{eqnarray}
\frac{\ell (\ell +1)}{2 \pi} C_{\ell}^{(\mathrm{VV})} &=& \frac{32}{9} f_{\mathrm{e}}^2(\omega) {\mathcal A}_{{\mathcal R}} \biggl(\frac{k_{0}}{k_{\mathrm{p}}}\biggr)^{n_{\mathrm{s}} -1} \, \ell^{n_{\mathrm{s}} -1} \int_{1}^{\infty} w^{n_{\mathrm{s}} - 5} \sqrt{w^2 -1} \, dw\,  \times 
\nonumber\\
&\times&\biggl[ \biggl(2 L_{{\mathcal R}}^2(w,\ell) e^{ - 2 (\ell/\ell_{\mathrm{t}})^2 w^2} + M_{{\mathcal R}}^2(w,\ell) c_{\mathrm{sb}} e^{- 2 (\ell/\ell_{\mathrm{S}})^2 w^2}\biggr) \nonumber\\
&+&   M_{{\mathcal R}}^2(w,\ell) c_{\mathrm{sb}} \, 
\cos{[2 (\gamma_{\mathrm{A}} w \ell)]}\, e^{- 2 (\ell/\ell_{\mathrm{S}})^2 w^2}
\nonumber\\
&+& 4 L_{{\mathcal R}}(w,\ell)  M_{{\mathcal R}}(w,\ell) \sqrt{c_{\mathrm{sb}}} \cos{(\gamma_{\mathrm{A}} w \ell)}
e^{-  [ (\ell/\ell_{\mathrm{S}})^2 +(\ell/\ell_{\mathrm{t}})^2] w^2}
\biggr],
\label{smallang}
\end{eqnarray}
where $c_{\mathrm{sb}} =[3 (R_{\mathrm{b}} +1)]^{-1/2}$ is 
the baryon sound speed. Following Ref. \cite{max2} we introduced in Eq. (\ref{smallang})
the quantities\footnote{It is understood that all the quantities are computed for 
at $z=z_{\mathrm{rec}}$; furthermore $r_{\mathrm{R}} = \rho_{\mathrm{R}}/\rho_{\mathrm{M}} = 4.15 \times 10^{-2} \omega_{\mathrm{M}} (z/1000)$.}  
\begin{eqnarray}
&&L_{{\mathcal R}}(w, \ell) = \alpha_{{\mathcal R}} - \beta_{{\mathcal R}} \, \ln{[w q_{\ell}]}, \qquad M_{{\mathcal R}}(w,\ell) = \overline{\alpha}_{{\mathcal R}} + \overline{\beta}_{{\mathcal R}} \, \ln{[w q_{\ell}]},
\nonumber\\
&& q_{\ell} = \biggl( \frac{\ell}{200\, d_{\mathrm{A}}(z)}\biggr) \sqrt{\frac{r_{R}}{z +1}},
\nonumber\\
&& \alpha_{{\mathcal R}} = \frac{R_{\mathrm{b}} +1}{6}, \qquad \beta_{{\mathcal R}} = \frac{R_{\mathrm{b}}}{6},
\nonumber\\
&& \overline{\alpha}_{{\mathcal R}}= - \frac{6}{25} \ln{(96)},\qquad \overline{\beta}_{{\mathcal R}} = - \frac{6}{25}.
\label{DEF}
\end{eqnarray}
In Eq. (\ref{DEF}) the quantity $d_{\mathrm{A}}(z)$ is 
related to the angular diameter distance $D_{\mathrm{A}}(z)$ as 
$d_{\mathrm{A}}(z) = \sqrt{\Omega_{\mathrm{M}0}} H_{0} D_{\mathrm{A}}(z)/2$. Furthermore $\gamma_{\mathrm{A}}$ and 
$\ell_{\mathrm{D}}$ can be estimated as follows\footnote{Following the usual convention we shall denote $\omega_{\mathrm{X}} = h_{0}^2 \Omega_{\mathrm{X}0}$ for a generic species.}
\begin{eqnarray}
\gamma_{\mathrm{A}} 
&=& \frac{d_{\mathrm{A}}^{-1}(z)}{\sqrt{3 R_{\mathrm{b}} ( z+1)}} \ln{\biggl[ \frac{\sqrt{1 +  R_{\mathrm{b}}} + 
\sqrt{R_{\mathrm{b}}}\sqrt{1 + r_{\mathrm{R}}}}{1 + \sqrt{r_{\mathrm{R}}\,R_{\mathrm{b}}}}\biggr]},
\nonumber\\
\ell_{\mathrm{D}} &=& k_{\mathrm{D}}\, D_{\mathrm{A}}(z)= 
\frac{2240 \, d_{\mathrm{A}}(z_{*})}{\sqrt{\sqrt{r_{\mathrm{R}} +1} - \sqrt{r_{\mathrm{R}}}}} 
\biggl(\frac{z}{10^{3}} \biggr)^{5/4} \, \omega_{\mathrm{b}}^{0.24} \omega_{\mathrm{M}}^{-0.11}.
\end{eqnarray}
The Silk multipole is just expressed in terms 
of $\ell_{\mathrm{t}}$ and $\ell_{\mathrm{D}}$ as $\ell_{\mathrm{S}} = \ell_{\mathrm{t}} \ell_{\mathrm{D}}/\sqrt{\ell_{\mathrm{t}}^2 + \ell_{\mathrm{D}}^2}$. With the same approach it is possible to express, for small angular scales, the 
cross-correlation $C_{\ell}^{(\mathrm{VT})}(\omega)$.

\begin{figure}[!ht]
\centering
\includegraphics[height=6cm]{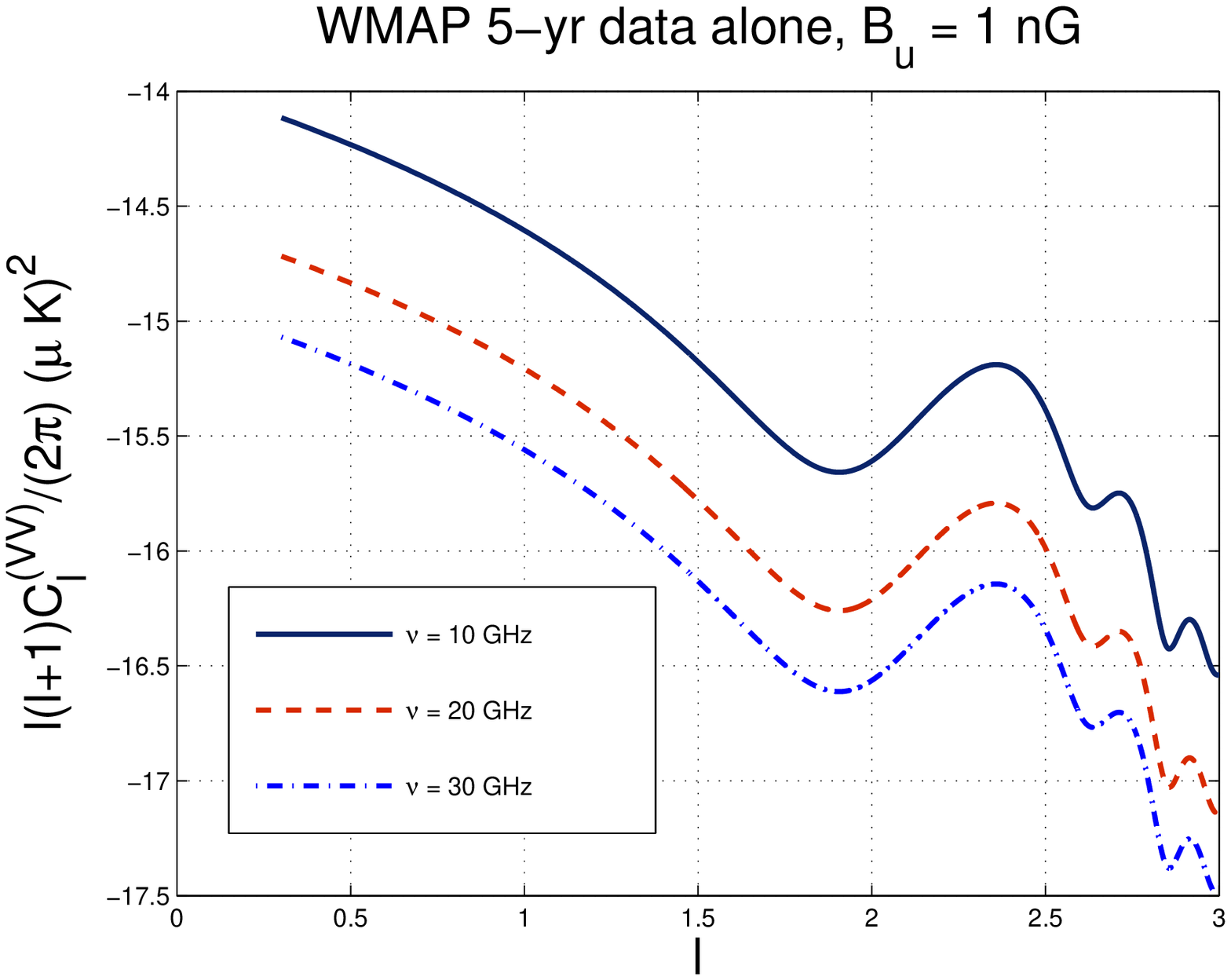}
\includegraphics[height=6cm]{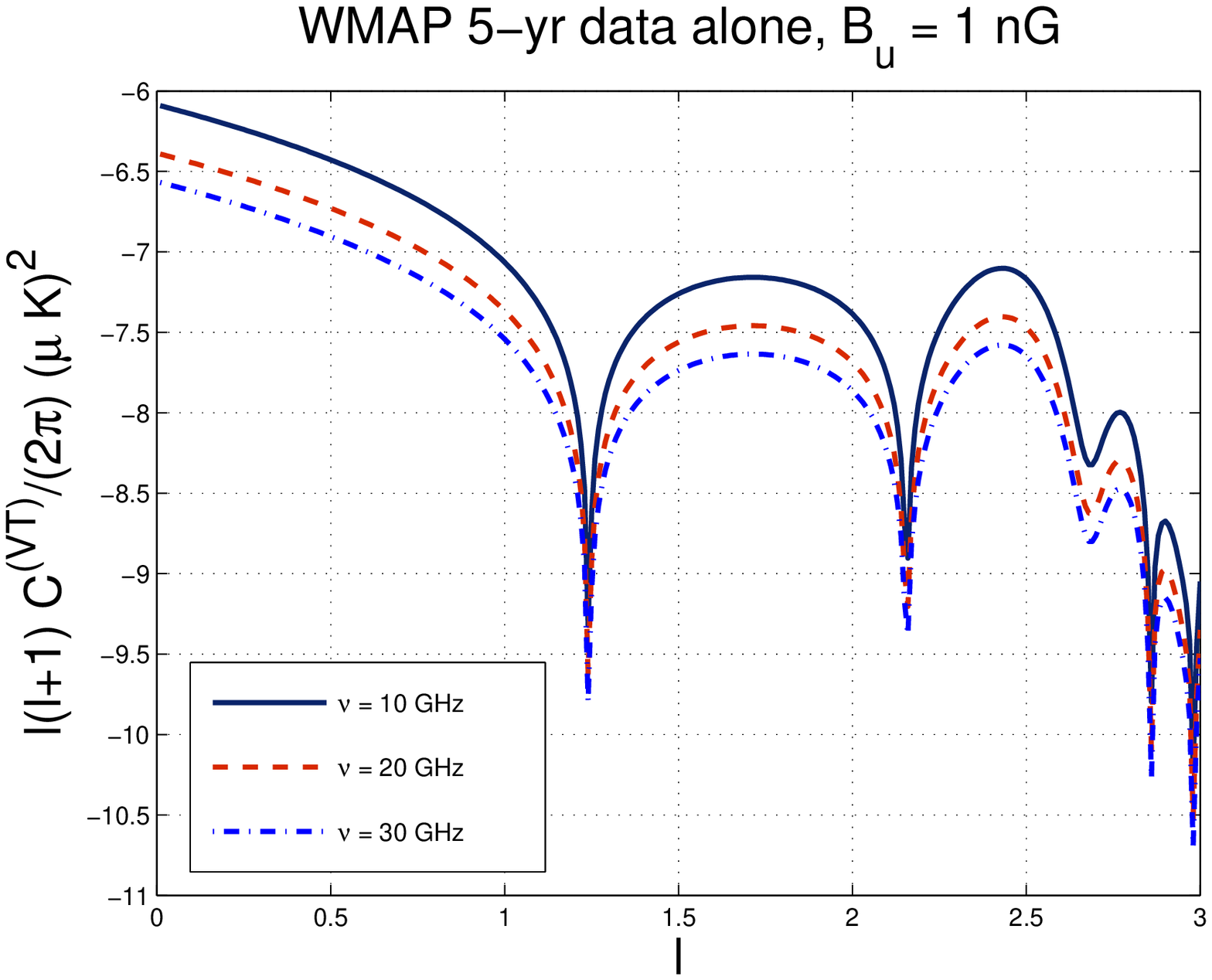}
\caption[a]{The VV and VT angular power spectra are illustrated for small angular scales and for different values of the frequency and for fixed magnetic field intensity. As in Fig. \ref{figure2} a double logarithmic scale 
has been employed in both plots.}
\label{figure3}      
\end{figure}

\begin{figure}[!ht]
\centering
\includegraphics[height=6cm]{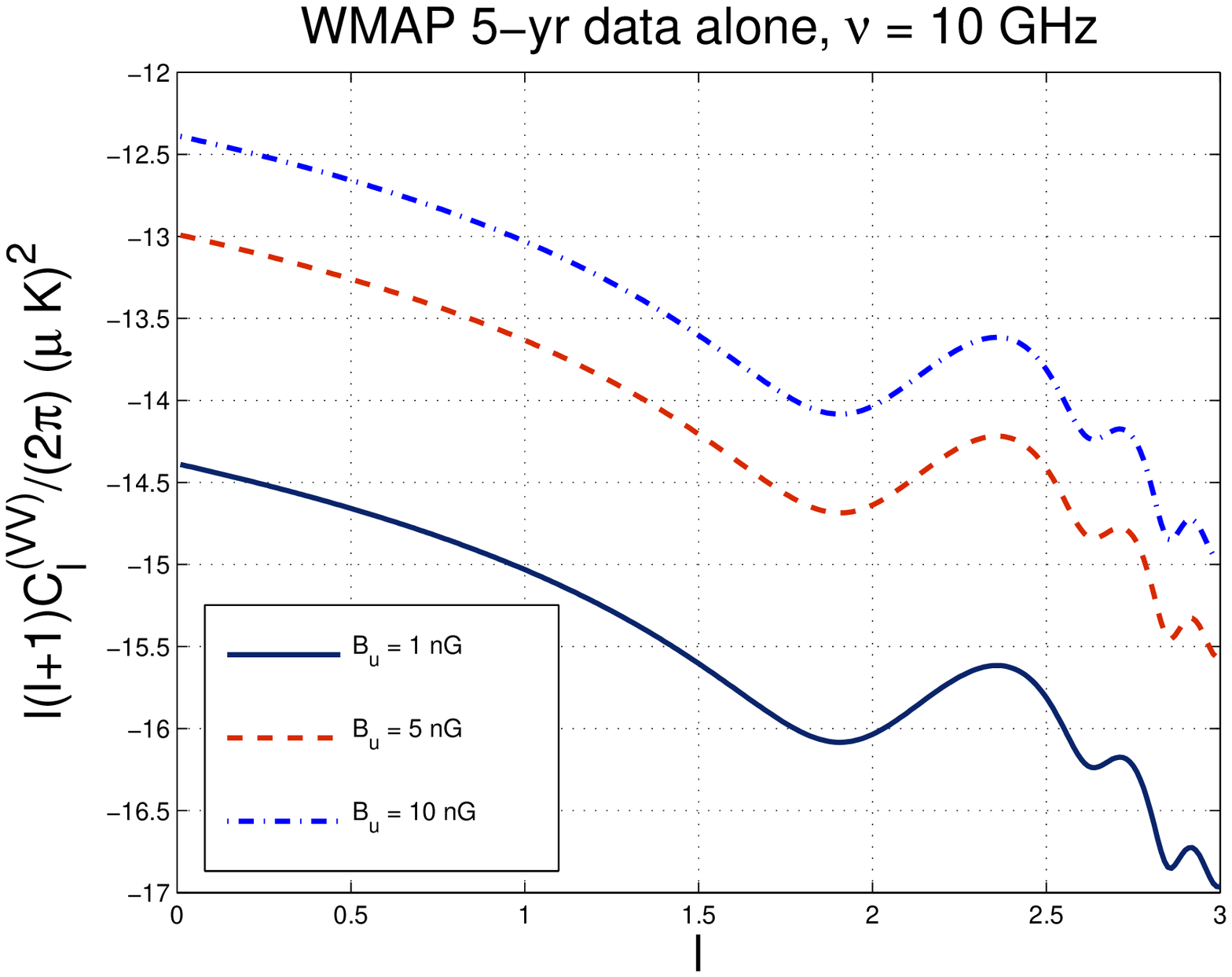}
\includegraphics[height=6cm]{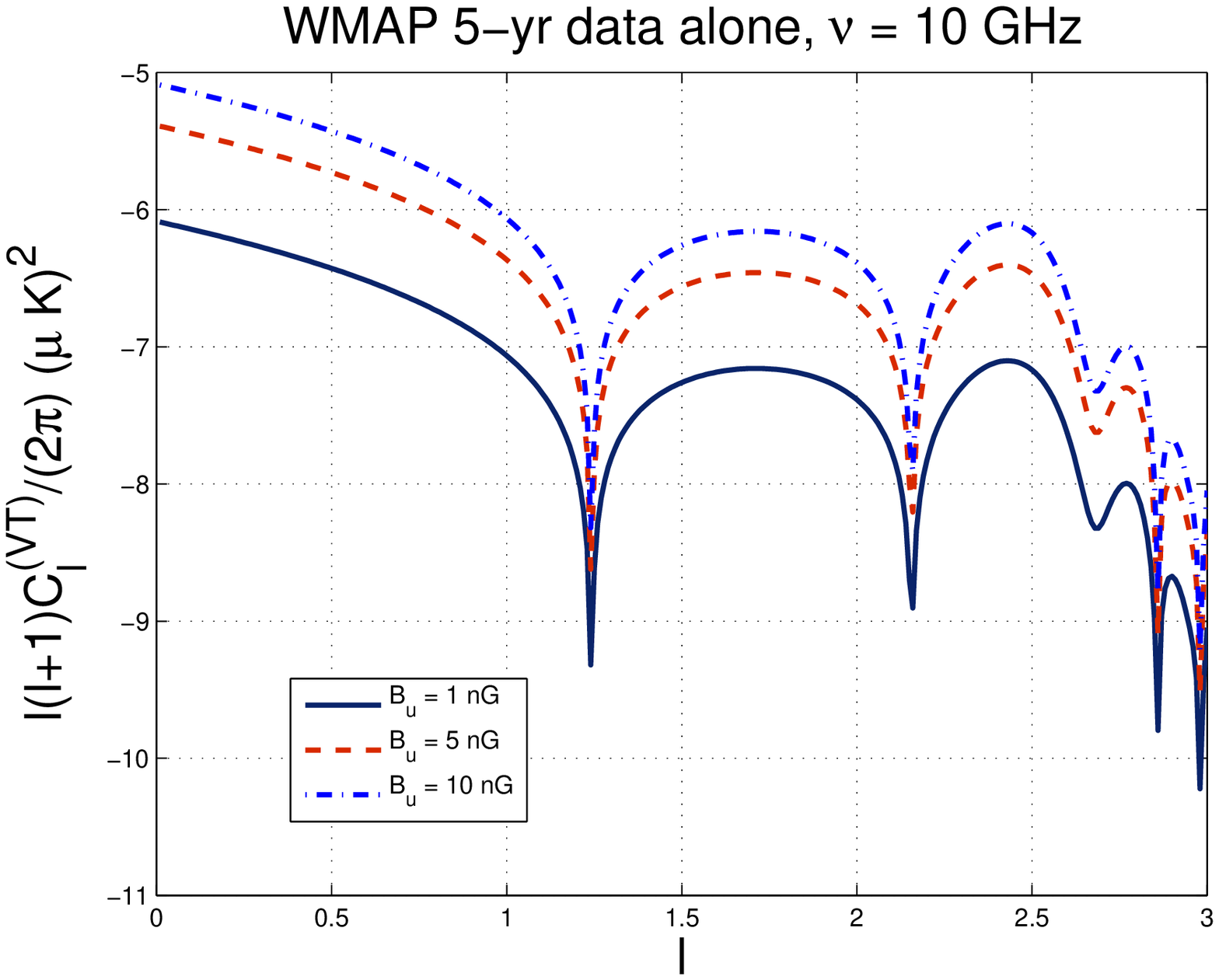}
\caption[a]{The VV and VT angular power spectra are illustrated for small angular scales and for different values of the magnetic field at a fixed value of the frequency. 
As in the previous figures, a double logarithmic scale has been employed.}
\label{figure4}      
\end{figure}

For more reliable estimates at small scales fully numerical methods should be employed and 
the results, consistent with the previous analytic estimates,  are illustrated in Fig. \ref{figure3} and \ref{figure4}. 
In Fig. \ref{figure3} the VV and VT angular power spectra are illustrated 
on a double logarithmic scale. The spikes appearing in the VT correlation 
are the usual feature displayed when plotting the modulus of the cross-correlation.
The same spikes occur when plotting the logarithm of the modulus of the TE power spectrum.  In Fig. \ref{figure3} 
the magnetic field is fixed to $1$ nG while the frequency of the channel 
is allowed to vary.  In Fig. \ref{figure4} the VV and VT correlations are illustrated for a fixed value of the 
frequency but for various magnetic field intensities. 

The moment has now come to compare the signal from circular dichroism with the signals expected from the linear polarization. 
The spirit of the forthcoming considerations is just to compare the orders of magnitude of the different contributions.
While this step is mandatory within the present analysis it is also not conclusive, from the experimental 
point of view. Indeed different signals and different correlation functions inherit 
from nature different systematic effects. The latter problem is 
of course extremely important and will not be treated here.

Let us first of all compare the V-mode signal with the measurements (and expectations) of the $\Lambda$CDM paradigm.
A simple comparison is illustrated in Tab. \ref{TABLE1} where the different angular power spectra are reported at the 
peak. The TT power spectrum at the first acoustic peak is of the order of $6000\,\,(\mu \mathrm{K})^2$. The TE 
and the EE angular power spectra peak for larger multipoles.  The WMAP team measured with reasonable 
accuracy the region of the first anticorrelation peak of the TE power spectrum and the lowest 
multipoles of the EE power spectrum. The recent QUAD measurements gave a rather 
interesting evidence of the oscillations in the EE angular power spectra at larger $\ell$. The absolute 
value of the TE angular power spectrum peaks around $\ell \sim 750$ and it is of the order of 
$130\,\,(\mu \mathrm{K})^2$. The EE correlation  reaches a value of  $40\,\,(\mu \mathrm{K})^2$
for $\ell \sim 1000$. The quoted figures are consistent with the expectations of the $\Lambda$CDM 
scenario with no tensors (also sometimes called vanilla $\Lambda$CDM).  
\begin{table}[!ht]
\begin{center}
\begin{tabular}{||l|c|c|c|c|c|c||}
\hline
Data & $\ell_{\mathrm{peak}}$ & $\ell_{\mathrm{peak}} (\ell_{\mathrm{peak}} +1) C^{(\mathrm{XX})}_{\ell_{\mathrm{peak}}}/(2\pi)$\\
\hline
TT &   $220$  & $5756\,\, (\mu \mathrm{K})^2$  \\
EE & $1000$ & $    40\,\,  (\mu \mathrm{K})^2$ \\
TE &$ 750 $ & $130\,\,  (\mu \mathrm{K})^2$\\
VT & $\ell < 50$ & $10^{-6}\,\,  (\mu \mathrm{K})^2$  \\
VV & $\ell <50$ &  $10^{-14}\,\,  (\mu \mathrm{K})^2$\\
\hline
\end{tabular}
\caption{The values of the different angular power spectra at the peak (llustrative figures for $B_{\mathrm{u}} = 1$nG and $\nu=10$GHz.}
\label{TABLE1}
\end{center}
\end{table}
In the vanilla $\Lambda$CDM the only potential source of B-mode polarization is represented by gravitational 
lensing of the primary anisotropies. The typical values of the induced BB angular power spectrum range between 
 $10^{-8} \,\,  (\mu \mathrm{K})^2$ and $10^{-5}$ for $\ell <50$.
In Tab. \ref{TABLE1} the expectations for the VV and VT angular power spectra are reported in the 
case of a hypothetical low frequency instrument sensitive to V-mode polarization operating 
in a band with central frequency of the order of $10$ GHz. The intensity of the (comoving) magnetic field 
has been taken to be $1$ nG. As previously discussed both analytically and numerically the VV and VT 
power spectra are larger for low multipoles. For larger multipoles, however, the effects 
of the thickness of the visibility function and of the diffusive damping come into play only for 
$\ell \sim 1000$. In this sense the range of multipoles highlighted in Tab. \ref{TABLE1} should be complemented 
with the results illustrated in Figs. \ref{figure1}, \ref{figure2}, \ref{figure3} and \ref{figure4}. 

The comparison summarized in Tab. \ref{TABLE1} does not contemplate 
the BB angular power spectrum stemming from the tensor modes 
of the geometry which is regarded as the main target of running experiments such as 
Planck.
\begin{table}[!ht]
\begin{center}
\begin{tabular}{||l|c|c|c|c|c||}
\hline
Data & $r_{\mathrm{T}}$ &$n_{\mathrm{s}}$ & $\Omega_{\Lambda}$&$\Omega_{\mathrm{M}0}$\\
\hline
 WMAP5 alone& $<0.43$ &$0.986 \pm 0.22$ & $0.770_{-0.032}^{+0.033}$&$0.230_{-0.033}^{0.032}$ \\
WMAP5 + Acbar& $< 0.40  $& $0.985_{-0.020}^{0.019}$&$0.767 \pm 0.032$&$0.233\pm 0.032$ \\
WMAP5+ LSS + SN &$<0.20$ &$0.968 \pm 0.015$&$0.725 \pm 0.015$&$0.275 \pm 0.015$\\
WMAP5+ CMB data & $<0.36$ & $0.979\pm 0.020$&$0.775\pm 0.032$&$0.225\pm 0.032$\\
\hline
\end{tabular}
\caption{The change in determination of the parameters of the tensor background for three different choices 
of cosmological data sets.}
\label{TABLE2}
\end{center}
\end{table}
The B-mode power spectrum induced by relic gravitons  peaks for $\ell \sim 90$ corresponding 
to an angular scale of roughly $2$ deg. The signal, however, depends upon $r_{\mathrm{T}}$ (i.e. the ratio 
between the tensor and the scalar power spectrum) 
for which only upper limits exist. Depending upon the data sets chosen for the analysis 
the putative limit on $r_{\mathrm{T}}$ slightly changes. The situation 
is quickly summarized in Tab. \ref{TABLE2} where the upper limits on $r_{\mathrm{T}}$ are 
reported at the pivot scale $k_{\mathrm{p}} =0.002\, \mathrm{Mpc}^{-1}$ and in the case 
where the scalar spectral index does not run. Slightly larger values of $r_{\mathrm{T}}$ 
are allowed if $n_{\mathrm{s}}$ is allowed to run but this aspect will not be essential 
for the present considerations \footnote{In the case of running the bounds on $r_{\mathrm{T}}$ range 
from $0.58$ (in the case of the WMAP 5-yr data alone) to $0.54$ if we combine the 
WMAP data with the large-scale structure data and with the supernova data.}. 
\begin{figure}[!ht]
\centering
\includegraphics[height=6cm]{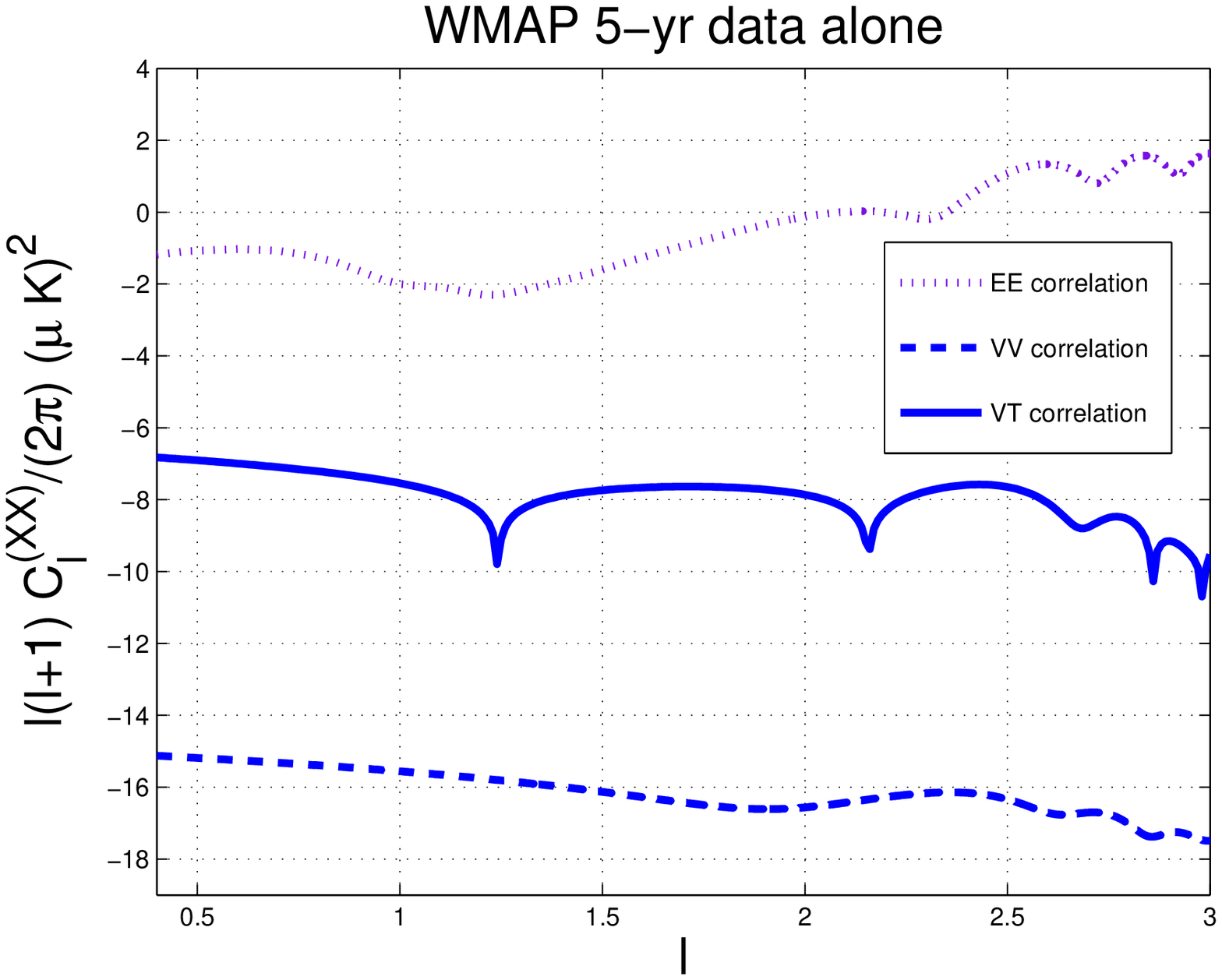}
\includegraphics[height=6cm]{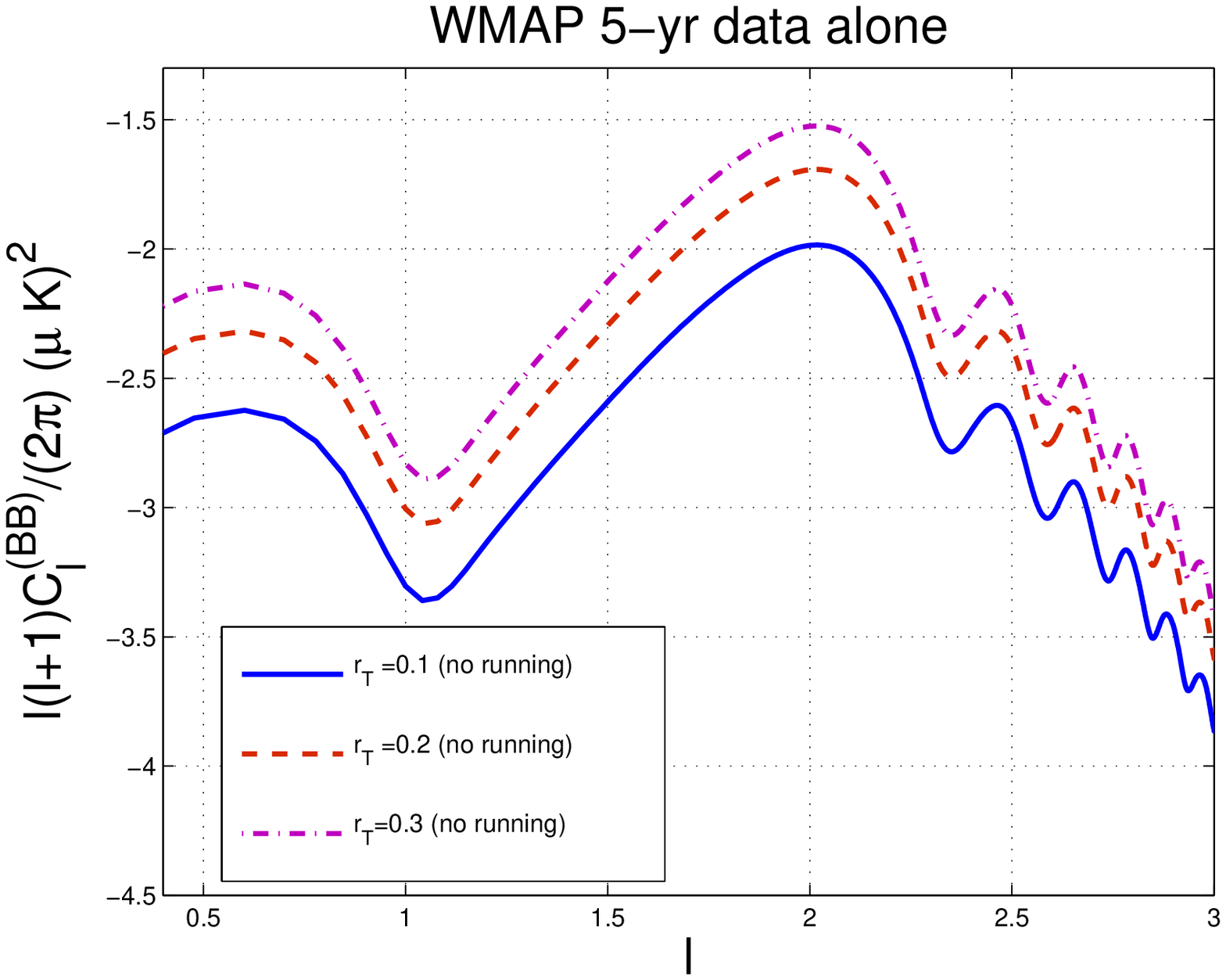}
\caption[a]{The VV and VT angular power spectra are compared with other polarization signals arising 
in the $\Lambda$CDM paradigm and in its neighboring extensions.}
\label{figure5}      
\end{figure}
The comparison between the VV and VT power spectra and the other polarization 
power spectra is also illustrated, more visually, in Fig. \ref{figure5}. 
In the plot at the left  of Fig. \ref{figure5} the upper curve corresponds, as indicated by the 
legend, to the EE angular power spectra obtained from the best fit to the WMAP 5-yr data 
alone (see Eq. (\ref{Par1})). In the plot at the right 
the B-mode polarization is illustrated when a tensor mode contribution is 
allowed.  

It is tempting to speculate, at this point, that, indeed, low frequency instruments 
could make the difference for scrutinizing a potential V-mode polarization. 
In this respect the results and the techniques of \cite{gs1,gs2,gs3} (as well as the earlier 
results of \cite{LF1,LF2}) could be probably revisited in the light of the considerations 
developed here. It has been shown that the VT correlation for a comoving magnetic field 
from $5$ to $10$ nG can be as large as $10^{-5}\, (\mu \mathrm{K})^2$ at $10$ GHz for $\ell < 20$ (i.e. 
large angular separations). This means that for frequencies ${\mathcal O}(\mathrm{MHz})$, the resulting signal 
could be even $6$ or $7$ orders of magnitude larger than a putative B-mode 
signal from gravitational lensing which is between $10^{-8}\, (\mu \mathrm{K})^2$ and $10^{-6}\, (\mu \mathrm{K})^2$. 

\renewcommand{\theequation}{5.\arabic{equation}}
\setcounter{equation}{0}
\section{Concluding considerations}
\label{sec5}

In the present paper it has been argued that the presence of a large-scale magnetic field 
prior to equality can affect the photon-electron and the photon-ion scattering. In this 
process the radiation becomes circularly polarized and the induced VV and VT 
angular power spectra have been computed. 
The analysis reported in the present paper has to be regarded as very preliminary.
Indeed the considerations reported here can be refined both at the theoretical as well as 
at the more observational level. At the same time this investigation certainly  
opens the way for a more direct use of the circular polarization as a specific 
diagnostic of pre-decoupling magnetism. The reported results might also be regarded as
a modest spur for those observers and experimenters 
whose aim is a direct measurement (or a plausible upper limit) on the circular dichroism 
of the Cosmic Microwave Background. 

\section*{Acknowledgment}

It is a pleasure to thank continued conversations and exchanges of ideas with G. Sironi, 
M. Gervasi, and A. Tartari.  Interesting conversations with N. Mandolesi are also 
acknowledged. The author wishes also to thank T. Basaglia and A. Gentil-Beccot of the CERN scientific information service  for swiftly providing copies of several references.

\newpage

\begin{appendix}
\renewcommand{\theequation}{A.\arabic{equation}}
\setcounter{equation}{0}
\section{Derivation of the scattering matrix}
\label{APPA}
In a conformally flat background geometry characterized by a metric tensor  $g_{\mu\nu}(\tau) = a^{2}(\tau) \eta_{\mu\nu}$ the electron-ion plasma can be described  by the well known set of two-fluid equations (see, e.g., \cite{max5,max4}):
\begin{eqnarray}
&& \vec{\nabla}\cdot \vec{E} = 4 \pi e (n_{\mathrm{i}} - n_{\mathrm{e}}), \qquad \vec{\nabla}\cdot\vec{B} =0
\label{Mx1}\\
&&\vec{\nabla} \times \vec{B} = 4 \pi e (n_{\mathrm{i}} \vec{v}_{\mathrm{i}} - n_{\mathrm{e}} \vec{v}_{\mathrm{e}}) + 
\frac{\partial \vec{E}}{\partial \tau},\qquad \vec{\nabla}\times \vec{E} = - \frac{\partial \vec{B}}{\partial\tau},
\label{Mx2}
\end{eqnarray}
where the electromagnetic fields as well as the concentrations of electrons and ions are comoving, i.e. 
\begin{equation}
\vec{B}(\vec{x},\tau) = a^2(\tau) \vec{{\mathcal B}}(\vec{x}, \tau),\qquad \vec{E}(\vec{x},\tau) = a^2(\tau)
 \vec{{\mathcal E}}(\vec{x}, \tau)\qquad n_{\mathrm{e,\,i}}(\vec{x},\tau) = a^3(\tau) \tilde{n}_{\mathrm{e,\,i}}(\vec{x},\tau).
 \end{equation}
The evolution equations of the electron and ion velocities can be written as 
\begin{eqnarray}
&& \frac{d \vec{v}_{\mathrm{e}}}{d\tau} + {\mathcal H} \vec{v}_{\mathrm{e}} = - \frac{e}{m_{\mathrm{e}}a} [ \vec{E} + \vec{v}_{\mathrm{e}} \times \vec{B}],
\label{electronv}\\
&& \frac{d \vec{v}_{\mathrm{i}}}{d\tau} + {\mathcal H} \vec{v}_{\mathrm{i}} =  \frac{e}{m_{\mathrm{p}}a} [ \vec{E} + \vec{v}_{\mathrm{i}} \times \vec{B}],
\label{ionv}
\end{eqnarray}
where $m_{\mathrm{e}}$ and $m_{\mathrm{p}}$ are, respectively, 
the electron and the ion masses; the velocities are related, as usual, to the comoving 
three-momentum $\vec{v}_{\mathrm{e,\,i}}= \vec{q}_{\mathrm{e,\,i}}/\sqrt{q_{\mathrm{e,\,i}}^2 + m_{\mathrm{e,\,p}}^2 a^2}$.
The explicit dependence upon the scale factor at the right-hand side of Eqs. (\ref{electronv}) and (\ref{ionv}) 
arises because the plasma is cold: both electrons and ions are non-relativistic and the mass dependence 
breaks explicitly the Weyl invariance of the whole system already at the level of the Vlasov-Landau 
equations for the distribution function \cite{max4,max5}. Finally, in Eqs. (\ref{electronv}) and (\ref{ionv})
${\mathcal H} = a'/a$ enters directly the Friedmann-Lema\^itre equations:
\begin{equation}
{\mathcal H}^2 = \frac{8 \pi G}{3} a^2\rho_{\mathrm{t}}, \qquad {\mathcal H}^2 - {\mathcal H}' = 4 \pi G a^2 (p_{\mathrm{t}} + 
\rho_{\mathrm{t}}),
\label{FL}
\end{equation}
where $\rho_{\mathrm{t}}$ and $p_{\mathrm{t}}$ denote the total energy density and the total pressure while 
the prime stands for a derivation with respect to the conformal time coordinate $\tau$. Since the electron and ion 
concentrations are comoving, they simply obey the following pair of equations
\begin{equation}
n_{\mathrm{e}}'  + \vec{\nabla}\cdot(n_{\mathrm{e}} \vec{v}_{\mathrm{e}}) =0,\qquad 
n_{\mathrm{i}}'  + \vec{\nabla}\cdot(n_{\mathrm{i}} \vec{v}_{\mathrm{i}}) =0.
\label{conc}
\end{equation}
Further details on the description of globally neutral plasma in Friedmann-Robertson-Walker backgrounds can be 
found, for instance, in \cite{rev2,max1,max2,max4} (see also \cite{max6,max7} for earlier results).

The purpose is now to derive the components of the scattering matrix 
${\mathcal S}_{ij}(\omega,\mu,\varphi,\mu',\varphi')$ introduced in Eq. (\ref{SM}).
In the dipole approximation  \cite{jack}, the outgoing electric field can be written as:
\begin{equation}
\vec{E}^{\mathrm{out}}(\vec{x},\tau)  = \frac{1}{r n_{0}} \biggl[ \hat{r} \times \hat{r} \times \frac{d \vec{J}}{d \tau}\biggr],
\label{co4}
\end{equation}
where 
\begin{equation}
\vec{J}(\vec{x},\tau) = e (n_{\mathrm{i}} \vec{v}_{\mathrm{i}} - n_{\mathrm{e}}\vec{v}_{\mathrm{e}})  
= e n_{0}( \vec{v}_{\mathrm{i}} - \vec{v}_{\mathrm{e}});
\label{co4a}
\end{equation}
the quantity $n_{0}= n_{\mathrm{e}} = n_{\mathrm{i}}$ denotes the common value of the (comoving) electron and ion concentrations.
In components the outgoing electric field can be written as 
\begin{equation}
E_{k} =  \frac{1}{n_{0} r}[(\vec{J}^{\,\prime}\cdot \hat{r}) \hat{r}_{k} - J^{\prime}_{k}],
\label{co6}
\end{equation}
where the prime denotes, as usual, a derivation with respect to the conformal time coordinate. 
The maximum of the microwave background arises today for typical photon energies of the order of $10^{-3}$ 
eV corresponding to a typical wavelength of the mm. At the time of decoupling  (i.e. $z_{\mathrm{dec}} \simeq 1089$ according to \cite{WMAP5a,WMAP5b,WMAP5c}) the wavelength of the radiation was of the order 
of $10^{-3} \mathrm{mm} \simeq \mu\mathrm{m}$.  Since the magnetic field 
we are interested in is inhomogeneous on a much larger length scale we can use the guiding 
centre approximation \cite{gc1} stipulating that 
\begin{equation}
B_{i}(\vec{x},\tau) \simeq B_{i}(\vec{x}_{0}, \tau)  + (x^{j} - x_{0}^{j}) \partial_{j} B_{i} +...
\label{co6a}
\end{equation}
where the ellipses stand for the higher orders in the gradients leading, both, to curvature and drift corrections
which will be neglected throughout. Fixing a local coordinate system with three orthogonal axes 
$\hat{x}$, $\hat{y}$ and $\hat{z}$ the components of the accelerations for the electrons are
\begin{eqnarray}
&&a^{(\mathrm{e})}_{x} = - \frac{\omega_{\mathrm{p\,\,e}}^2}{4\pi n_{0} e [ 1 - f_{\mathrm{e}}^2(\omega)]} [ E_{x} - i f_{\mathrm{e}}(\omega) E_{y}],~~~~
a^{(\mathrm{i})}_{x} = \frac{\omega_{\mathrm{p\,\,i}}^2}{4 \pi n_{0} e [ 1 - f_{\mathrm{i}}^2(\omega)]} [ E_{x} + i f_{\mathrm{i}}(\omega) E_{y}],
\nonumber\\
&&a^{(\mathrm{e})}_{y} = 
-\frac{\omega_{\mathrm{p\,\,e}}^2}{4 \pi n_{0} e [ 1 - f_{\mathrm{e}}^2(\omega)]}  [ E_{y} + i f_{\mathrm{e}}(\omega) E_{x}],~~~~
a^{(\mathrm{i})}_{y} =\frac{\omega_{\mathrm{p\,\,i}}^2}{4\pi n_{0} e [ 1 - f_{\mathrm{i}}^2(\omega)]}  [ E_{y} - i f_{\mathrm{i}}(\omega) E_{x}],
\nonumber\\
&& a^{(\mathrm{e})}_{z} = - \frac{\omega_{\mathrm{p\,\,e}}^2}{4 \pi n_{0} e } E_{z},~~~~
a^{(\mathrm{i})}_{z} =\frac{\omega_{\mathrm{p\,\,i}}^2}{4\pi n_{0} e} E_{z}
\label{co9}
\end{eqnarray}
where $f_{\mathrm{e,\,\,i}} = \omega_{\mathrm{B\,\,e,\,\,i}}/\omega$; $\omega_{\mathrm{B\,\,e,\,\,i}}$ and 
 $\omega_{\mathrm{p\,\,e,\,\,i}}$ are, respectively, the Larmor and the plasma frequencies either of the electrons or of the ions (see Eqs. (\ref{LARM}) and \ref{PLAS})). 
The magnetic field is oriented along the $z$ axis and only the lowest order in the gradient expansion is kept.
Of course, as it is well known, higher order 
will induce both gradient drifts as well as curvature drifts (see, e.g. \cite{gc1}). We consider these terms to be negligible in the first approximation. 
Here we are interested in the scattering of electrons and photons not in the stellar atmosphere but rather at the decoupling time when the physical wavelength 
of the photons is minute in comparison with the inhomogeneity scale of the magnetic field which is of the order of the Hubble radius and even larger \cite{max1,max2}. The guiding centre approximation is pretty robust as far as the magnetized 
scattering in concerned. For sufficiently small angular scales (i.e. $\ell \gg 100$) the radial direction, in spherical 
coordinates, coincide (approximately) with the $\hat{z}$ direction. For this reason, in some related paper 
the modulus of the magnetic field has been taken as $ B = \hat{z} \cdot \vec{B} \simeq \hat{n}\cdot \vec{B}$
where the last approximate equality follows in the limit  of small angular scales. In the limit $\ell \gg 100$ 
the two-sphere actually degenerate into a plane and this is the reason why one can trade the spherical decomposition 
for the plane wave expansion.
 
It is also possible to write the difference of the accelerations of electrons and ions, namely, 
$\vec{A} = (\vec{a}^{(\mathrm{e})} - \vec{a}^{(\mathrm{i})})$ whose components in the local frame are 
\begin{eqnarray}
&& A_{x} = - \frac{\omega_{\mathrm{p\,\,e}}^2}{4\pi n_{0} e [ 1 - f_{\mathrm{e}}^2(\omega)]} [ \Lambda_{1}(\omega)
 E_{x} - i f_{\mathrm{e}}(\omega) \Lambda_{2}(\omega) E_{y}],
\label{co17}\\
&& A_{y}  = - \frac{\omega_{\mathrm{p\,\,e}}^2}{4\pi n_{0} e [ 1 - f_{\mathrm{e}}^2(\omega)]} [ \Lambda_{1}(\omega) E_{y} + i f_{\mathrm{e}}(\omega) \Lambda_{2}(\omega) E_{x}],
\label{co18}\\
&& A_{z} = -  \frac{\omega_{\mathrm{p\,\,e}}^2}{4\pi n_{0} e}\Lambda_{3}(\omega) E_{z},
\label{co19}
\end{eqnarray}
where $\Lambda_{1}(\omega)$, $\Lambda_{2}(\omega)$ and $\Lambda_{3}(\omega)$ have been already 
defined in Eqs. (\ref{LAM1}), (\ref{LAM2}) and (\ref{LAM3}). 
To compute the elements of the scattering matrix it is preferable to pass from the Cartesian components 
of the incident electric fields to the polar components; the relation between the fields in the two basis is 
given by:
\begin{eqnarray}
&& E_{x} = E_{\vartheta}'\cos{\vartheta'} \cos{\varphi'} - \sin{\varphi'} E_{\varphi}',
\label{co27}\\
&&  E_{y} = E_{\vartheta}'\cos{\vartheta'} \sin{\varphi'} + \cos{\varphi'} E_{\varphi}',
\label{co28}\\
&& E_{z} = - \sin{\vartheta'} E_{\vartheta}'.
\label{co29}
\end{eqnarray}
To avoid a proliferation of superscripts in the intermediate expressions the 
ingoing electric fields $ E_{\vartheta}'$ and $E_{\varphi}'$ are renamed as 
$E_{1}$ and $E_{2}$, i.e.
\begin{equation}
E_{1}= E_{\vartheta}', \qquad E_{2}= E_{\varphi}'.
\label{co30}
\end{equation}
Using Eq. (\ref{co9}) the scattered electric field, in the dipole approximation, 
become
\begin{equation}
\vec{E} = E_{\vartheta} \hat{\vartheta} + E_{\varphi} \hat{\varphi},
\label{co31}
\end{equation}
where recalling the notations of Eqs. (\ref{LAM1})--(\ref{zetaomega})
\begin{eqnarray}
E_{\vartheta}(\omega,\mu,\mu',\varphi,\varphi') &=& \frac{r_{\mathrm{e}}}{r} \biggl\{ \biggl[ \Lambda_{1}(\omega) \zeta(\omega)\, \mu \biggl( \mu' E_{1} \cos{(\varphi' -\varphi)} - 
E_{2} \sin{(\varphi' - \varphi)}\biggr)
\nonumber\\
&-& \sqrt{1 - \mu^2} \sqrt{1 - {\mu'}^2} \Lambda_{3}(\omega) E_{1} \biggr]
\nonumber\\
&-& i \Lambda_{2}(\omega) f_{\mathrm{e}}(\omega) \zeta(\omega) \mu \biggl[ \mu' E_{1} \sin{(\varphi' - \varphi)} + E_{2} \cos{(\varphi'- \varphi)}\biggr]
\biggr\},
\label{co35}\\
E_{\varphi}(\omega,\mu,\mu',\varphi,\varphi') &=& \frac{r_{\mathrm{e}}}{r} \biggl\{ \zeta(\omega) \Lambda_{1}(\omega)
 \biggl[ \mu' E_{1} \sin{(\varphi' - \varphi)} + E_{2} \cos{(\varphi' - \varphi)}\biggr] 
\nonumber\\
&+& i f_{\mathrm{e}}(\omega) \Lambda_{2}(\omega) \zeta(\omega) \biggl[ \mu' E_{1} \cos{(\varphi' - \varphi)} - E_{2} \sin{(\varphi' - \varphi)}
\biggr]\biggr\}.
\label{co36}
\end{eqnarray}
Thus, collecting the different factors and rearranging the final expressions we shall have that the scattered 
electric fields can be written as 
\begin{equation}
E_{\vartheta}(\omega,\mu,\mu',\varphi,\varphi')  = \frac{r_{\mathrm{e}}}{r} [ A \, E_{1} + B\, E_{2}],
\qquad
E_{\varphi}(\omega,\mu,\mu',\varphi,\varphi')  = \frac{r_{\mathrm{e}}}{r} [ C \, E_{1} + D\, E_{2}],
\end{equation}
where 
\begin{eqnarray}
A(\omega,\mu,\mu',\varphi,\varphi') &=& \zeta(\omega) \mu \mu' \Lambda_{1}(\omega) \cos{(\varphi'- \varphi)} - \sqrt{1 - \mu^2} \sqrt{1 - {\mu'}^2} 
\Lambda_{3}(\omega)
\nonumber\\
&-& i \Lambda_{2}(\omega) f_{\mathrm{e}}(\omega) \zeta(\omega) \mu\mu' \sin{(\varphi'-\varphi)}
\label{A}\\
B(\omega,\mu,\mu',\varphi,\varphi') &=& - \zeta(\omega) \mu \Lambda_{1}(\omega) \sin{(\varphi'-\varphi)} - i \Lambda_{2}(\omega) f_{\mathrm{e}}(\omega)
 \zeta(\omega) \mu \cos{(\varphi' - \varphi)},
\label{B}\\
C(\omega,\mu,\mu',\varphi,\varphi') &=& \mu' \zeta(\omega) \Lambda_{1}(\omega)
 \sin{(\varphi' - \varphi)} + i f_{\mathrm{e}}(\omega) \Lambda_{2}(\omega) \zeta(\omega) \mu' \cos{(\varphi' - \varphi)},
\label{C}\\
D(\omega,\mu,\mu',\varphi,\varphi') &=&  \zeta(\omega) \Lambda_{1}(\omega) \cos{(\varphi' - \varphi)} - i f_{\mathrm{e}}(\omega)
 \Lambda_{2}(\omega) 
\zeta(\omega) \sin{(\varphi' - \varphi)}.
\label{D}
\end{eqnarray}
The Stokes parameters of the scattered radiation can be related 
to the Stokes parameters of the incident radiation in terms 
of the appropriate scattering matrix ${\mathcal S}$. As already mentioned in connection 
with Eq. (\ref{SM}), arranging the outgoing and the ingoing Stokes parameters in a pair of column vectors
\begin{equation}
{\mathcal I}_{\mathrm{out}} = ( I_{\vartheta}, \, I_{\varphi},\, U,\, V), \qquad 
{\mathcal I}_{\mathrm{in}} = (I_{1},\, I_{2},\, U',\, V'),
\end{equation}
the outgoing Stokes parameters are given as 
${\mathcal I}_{\mathrm{out}}= {\mathcal S} \, {\mathcal I}_{\mathrm{in}}$, 
where the various components of ${\mathcal S}$ are:
\begin{eqnarray}
&& {\mathcal S}_{11} = \frac{r_{\mathrm{e}}^2}{r^2} |A|^2,\qquad {\mathcal S}_{12}  =  \frac{r_{\mathrm{e}}^2}{r^2} |B|^2, 
\nonumber\\
&& {\mathcal S}_{13}  =  \frac{r_{\mathrm{e}}^2}{2 r^2} (A^{*} B + B^{*} A),\qquad 
{\mathcal S}_{14}  = i\, \frac{r_{\mathrm{e}}^2}{2 r^2} (A^{*} B - B^{*} A),
\label{row1}\\
&& {\mathcal S}_{21}  =  \frac{r_{\mathrm{e}}^2}{r^2} |C|^2,\qquad {\mathcal S}_{22}  =
  \frac{r_{\mathrm{e}}^2}{r^2} |D|^2,
\nonumber\\
&& {\mathcal S}_{23}  = \frac{r_{\mathrm{e}}^2}{2 r^2} (C^{*} D + D^{*} C)
  ,\qquad  {\mathcal S}_{24}  = i \frac{r_{\mathrm{e}}^2}{2 r^2} (C^{*} D - D^{*} C),
\label{row2}\\
&& {\mathcal S}_{31}  = \frac{r_{\mathrm{e}}^2}{r^2}(A^{*} C + A C^{*}),\qquad {\mathcal S}_{32} =  
\frac{r_{\mathrm{e}}^2}{r^2} (B^{*} D + D^{*} B),
\nonumber\\
&& {\mathcal S}_{33} = \frac{r_{\mathrm{e}}^2}{2 r^2} ( A^{*} D + A D^{*} + B C^{*} + B^{*} C),
\nonumber\\
&& {\mathcal S}_{34} = \frac{i \,r_{\mathrm{e}}^2}{2\, r^2} ( A^{*} D - A D^{*} + B C^{*} - B^{*} C),
\label{row3}\\
&& {\mathcal S}_{41} =   \frac{i \, r_{\mathrm{e}}^2}{r^2} (A C^{*}\,-\,A^{*} C),\qquad  
{\mathcal S}_{42} =  \frac{i\, r_{\mathrm{e}}^2}{r^2} (B D^{*}-B^{*} D), \qquad  
\nonumber\\
&& {\mathcal S}_{43} =  \frac{i \, r_{\mathrm{e}}^2}{2\,r^2} (A D^{*} - A^{*} D + B C^{*} - B^{*} C),  
\nonumber\\
&& {\mathcal S}_{44} =  \frac{r_{\mathrm{e}}^2}{2\,r^2} (A^{*} D+ A D^{*} - B^{*} C - B C^{*}).
\label{row4}
\end{eqnarray}
Using Eqs. (\ref{A})--(\ref{D}) inside Eqs. (\ref{row1})--(\ref{row4}) the 
various matrix elements can be readily obtained in explicit terms and 
have been reported from Eq. (\ref{P11}) to Eq. (\ref{P44}). 
\renewcommand{\theequation}{B.\arabic{equation}}
\setcounter{equation}{0}
\section{Details on the derivation of the source terms}
\label{APPB}
It is appropriate to give few details on the derivation of the source terms 
reported in Eqs. (\ref{CIF}), (\ref{CQF}) and (\ref{CVF}). 
Using the results of Eq. (\ref{INTS}) the source terms of Eqs. (\ref{Br1})--(\ref{Br4}) can be rewritten in explicit terms as
\begin{eqnarray}
C_{\mathrm{I}}(\omega, \mu) &=&
 \frac{3}{16} \biggl\{ 2 \Lambda_{3}(\omega)(1 - \mu^2) {\mathcal Z}_{1}
+ \zeta^2(\omega) \biggl[  \Lambda_{1}^2(\omega) + 
f_{\mathrm{e}}^2(\omega) \Lambda_{2}^2(\omega)\biggr] (1 + \mu^2) 
{\mathcal Z}_{2} 
\nonumber\\
&+& \biggl[ 2 \Lambda_{3}(\omega) ( 1 -\mu^2)- \zeta^2(\omega) \biggl( \Lambda_{1}^2(\omega) + f_{\mathrm{e}}^2(\omega) \Lambda_{2}^2(\omega)\biggl)(1 + \mu^2)\biggr] {\mathcal Z}_{3} 
\nonumber\\
&+&
4 f_{\mathrm{e}}(\omega) \zeta^2(\omega) \Lambda_{1}(\omega) \Lambda_{2}(\omega) (\mu^2 + 1) {\mathcal Z}_{4}\biggr\},
\nonumber\\
C_{\mathrm{Q}}(\omega, \mu) &=& \frac{3}{16} \biggl\{2 \Lambda_{3}(\omega) (1 - \mu^2) {\mathcal Z}_{1}
- \zeta^2(\omega) \biggl( \Lambda_{1}^2(\omega)+ f_{\mathrm{e}}^2(\omega) \Lambda_{2}^2(\omega) \biggr)
(1 - \mu^2){\mathcal Z}_{2}
\nonumber\\
&+& \biggl[2 \Lambda_{3}(\omega) + \zeta^2(\omega) \biggl(\Lambda_{1}^2(\omega)  + 
f_{\mathrm{e}}^2(\omega) \Lambda_{2}^2(\omega)\biggr) \biggr](1 - \mu^2){\mathcal Z}_{3}
\nonumber\\
&+& 4 f_{\mathrm{e}}(\omega) \zeta^2(\omega) \Lambda_{2}(\omega) \Lambda_{1}(\omega)(\mu^2 - 1) {\mathcal Z}_{4}\biggr\}.
\nonumber\\
C_{\mathrm{V}}(\omega,\mu) &=& \frac{3}{8} \biggl\{ \mu f_{\mathrm{e}} \zeta^2(\omega) 
\Lambda_{2}(\omega)\Lambda_{1}(\omega)  \biggl[{\mathcal Z}_{2} - {\mathcal Z}_{3} \biggr]
\nonumber\\
&+&  \mu \,\zeta^2(\omega) \biggl(\Lambda_{1}^2(\omega) + f_{\mathrm{e}}^2(\omega) \Lambda_{2}^2(\omega)\biggr)
{\mathcal Z}_{4}  \biggr\}.
\label{interm1}
\end{eqnarray}
where the four quantities ${\mathcal Z}_{i}$ with $i= 1,\,2,\,3,\,4$ 
denote the integral of the various brightness perturbations, i.e. 
\begin{eqnarray}
&& {\mathcal Z}_{1} = \int_{-1}^{1} ( 1 - {\mu'}^2)\, \Delta_{\mathrm{I}} \,
d\mu' = \frac{4}{3} [ \Delta_{\mathrm{I}0} + \Delta_{\mathrm{I}2}],
\nonumber\\
&& {\mathcal Z}_{2} = \int_{-1}^{1} ( 1 + {\mu'}^2)\, \Delta_{\mathrm{I}} \,d\mu',
=\frac{8}{3} \Delta_{\mathrm{I}0} - \frac{4}{3} \Delta_{\mathrm{I}2},
\nonumber\\
&& {\mathcal Z}_{3} = \int_{-1}^{1} ( 1 - {\mu'}^2)\, \Delta_{\mathrm{Q}} 
d\mu' = \frac{4}{3} [ \Delta_{\mathrm{Q}0} + \Delta_{\mathrm{Q}2}],
\nonumber\\
&& {\mathcal Z}_{4} = \int_{-1}^{1} \mu'\, \Delta_{\mathrm{V}} \,d\mu' = - 2 \, i\, \Delta_{\mathrm{V}1}.
\label{integral1}
\end{eqnarray}
To derive Eq. (\ref{integral1}) the standard multipole expansion 
 for the brightness perturbations has been assumed, i.e. 
\begin{equation}
\Delta_{\mathrm{X}}(\hat{n},\tau) =\sum_{\ell} (-i)^{\ell}\, (2 \ell +1) \, 
\Delta_{\mathrm{X}\ell} P_{\ell}(\mu).
\label{integral2}
\end{equation} 
Using the explicit expressions of Eq. (\ref{integral1}) inside 
Eq. (\ref{interm1}), the expressions reported in Eqs. (\ref{CIF}), (\ref{CQF}) 
and (\ref{CVF}) are quickly recovered. 
\end{appendix}

\end{document}